\documentclass[11pt,a4paper]{elsarticle}
\usepackage{setspace}
\usepackage{ecrc_RIAI}
\usepackage{amsmath}
\usepackage[margin=1in]{geometry}
\usepackage[T1]{fontenc}
\usepackage{lscape}
\usepackage{longtable}
\usepackage{mathrsfs,dsfont}
\usepackage{graphicx}
\usepackage{eurosym}
\usepackage{tablefootnote}
\usepackage{footmisc}
\usepackage{pdflscape}
\usepackage{afterpage}
\usepackage{capt-of}
\usepackage{subfig}
\usepackage{listings}
\usepackage{adjustbox}
\usepackage{ulem}
\usepackage{booktabs}
 \usepackage[dvipsnames,table,xcdraw]{xcolor}
\usepackage{multirow}
\usepackage{bbm}
\usepackage{bm}
\usepackage{hyperref}
\usepackage{mathrsfs}
\usepackage{algorithm}
\usepackage{dsfont}
\usepackage{color}

\usepackage[utf8]{inputenc} 
\usepackage[T1]{fontenc}    
\usepackage{hyperref}       
\usepackage{url}            
\usepackage{booktabs}       
\usepackage{nicefrac}       
\usepackage[compact]{titlesec} 
\usepackage{enumitem}
\usepackage{mathtools}
\usepackage{amsthm}
\usepackage{bm} 
\usepackage{mathtools}
\usepackage{mathrsfs} 
\usepackage{tikz}
\usetikzlibrary{matrix,calc,shapes,arrows}

\usepackage{array}
\usepackage{adjustbox}
\usepackage{float} 
\usepackage{longtable}
\usepackage[english]{babel}
\usepackage{parskip}
\usepackage{setspace}
\usepackage{lineno,hyperref}
\usepackage{bm}  
\usepackage[autostyle, italian=guillemets]{csquotes}
\usepackage{graphicx} 
\usepackage{array}
\usepackage{soul}
\usepackage{makecell}
\usepackage{xcolor}

\usepackage{color}
\usepackage{helvet}         
\usepackage{courier}        
\usepackage{type1cm}        
\usepackage{makeidx}         %

\makeatletter
\def\BState{\State\hskip-\ALG@thistlm}
\makeatother

\theoremstyle{definition}
\def\bSig\bm{\Sigma}

\usetikzlibrary{shapes.multipart} 

\begin{document}

\begin{center}
\textmd{\LARGE{\bfseries{{Bootstrap Cointegration Tests in ARDL Models}}}}
\end{center}
\begin{center}
\large{Stefano Bertelli$^{\textcolor{Blue}{\text{1}}}$,
 \enspace
Gianmarco Vacca$^{\textcolor{Blue}{\textbf{2}}}$}, \enspace
Maria Grazia Zoia$^{{\textcolor{Blue}{\textbf{3}}}}$\\ \vspace{7mm}
\begin{small}
$^{\textcolor{Blue}{\textbf{1}}}$CRO Area, Internal Validation and Controls Department, Operational Risk and ICAAP Internal Systems, 
Intesa Sanpaolo, Milan, Italy. {\textcolor{Blue}{\textbf{{stefano.bertelli@intesasanpaolo.com}} }  }
\vspace{3mm}
\\$^{\textcolor{Blue}{\textbf{2}}}$Department of Economic Policy, Universit\`a Cattolica, Largo Gemelli 1, Milan 20123, Italy.\\Email:\textcolor{Blue}{\textbf{\href{mario.faliva@unicatt.it}} } {\textcolor{Blue}{\textbf{{gianmarco.vacca@unicatt.it}} }}
\vspace{3mm}
\\$^{\textcolor{Blue}{\textbf{3}}}$Department of Economic Policy, Univerist\`a Cattolica, Largo Gemelli, 1 Milan 20123, Italy. Corresponding Author. Tel:+390272342948; Fax: +390272342324; Email:\textcolor{Blue}{\textbf{\href{maria.zoia@unicatt.it}} } {\textcolor{Blue}{\textbf{{maria.zoia@unicatt.it}} }}

\end{small}
\end{center}
\vspace{10mm}

\begin{quote}
\begin{center}
\noindent {\bf Abstract} \end{center}
{The paper proposes a new bootstrap approach to the Pesaran, Shin and Smith's bound tests in a conditional equilibrium correction model with the aim to overcome some typical drawbacks of the latter, such as inconclusive inference and distortion in size. The bootstrap tests are worked out under several data generating processes, including degenerate cases.
Monte Carlo simulations confirm the better performance of the bootstrap tests with respect to bound ones and to the asymptotic F test on the independent variables of the ARDL model.
It is also proved that any inference carried out in misspecified models, such as unconditional ARDLs, may be misleading.
Empirical applications highlight the importance of employing the appropriate specification and provide definitive answers to the inconclusive inference of the bound tests when exploring the long-term equilibrium relationship between economic variables.
}
\end{quote}
\vspace{10mm}
\begin{quote}
\vspace{5mm}
\end{quote}

\newpage
\section{Introduction}\label{sec: intro}
Cointegration and equilibrium correction are fundamental concepts to understand short-run and long-run properties of economic data, as they provide an appropriate framework for testing economic hypotheses about growth and fluctuations.
In any empirical investigation, economic time series must be properly analyzed to ascertain their integration and cointegration properties. Such analyses allow an effective interpretation of economic results and suggest possible modeling strategies and specifications that are consistent with the data, while also reducing the risk of spurious regressions.  \citep[see][]{banerjee1993co,davidson1978econometric,rao1997cointegration}.\\
In recent literature, the cointegration approaches proposed by ~\citet{granger1981some}, ~\citet{engle1987co}, ~\citet{johansen1990maximum} and ~\citet{pesaran2001} have become the paramount solutions to the problem of determining long-run equilibrium relationships among economic variables.
In particular, the latter contribution, also known as the Autoregressive Distributed Lag (ARDL) approach to cointegration or bound testing, has become prominent in empirical research thanks to its applicability also in cases of mixed order integrated variables, albeit with integration not exceeding the first order.   
This approach has been extensively used, having several advantages with respect to traditional statistical methods for testing cointegration.  First, it evades the necessity of  pre-testing the variables, thus avoiding some common practices that may prevent finding cointegrating relationships, such as dropping variables or transforming them into stationary form ~\citep[see, for example][]{mcnown2018bootstrapping}.
Furthermore, cointegrating testing is  performed in an ARDL model that allows different lag orders for each variable \citep{McNown2017}, thus providing a more flexible framework than other commonly employed approaches. The original bound tests proposed by \citet{pesaran2001} are an $F$-test for the significance of the coefficients of all lagged level variables entering the error correction term, and a $t$-test for the coefficient of the lagged dependent variable. 
When either the dependent or the independent variables do not appear in the long-run relationship, a degenerate case arises. The bound $t$-test provides answers on the occurrence of a degenerate case of first type, while the occurrence of the other degenerate case can be verified by testing whether the integration order of the dependent variable is $I(1)$. Indeed, a stationary dependent variable would rule out the existence of a long-run relationship between the latter and the other variables of the model.
Recently \citet{mcnown2018bootstrapping} pointed out how, due to the low power problem of unit root tests, investigating the presence of second type degeneracy by testing the integration order of the dependent variable may lead to incorrect conclusions. Therefore, they suggested checking this type of degeneracy by verifying the significance of the lagged levels of the independent variables via an extra $F$-test. \\
\citet{pesaran2001} derived the asymptotic distributions of the bound tests under the null hypothesis of the absence of a long-run cointegrating relationship and used stochastic simulations to compute two sets of critical values: one for the case of stationary regressors and another for first-order integrated regressors.  The former set represents the upper bound for the acceptance of the null, the latter the lower bound for its rejection. 
The inference is inconclusive if the test statistic lies between the said bounds, unless the integration order of the regressors is known. The asymptotic distributions of the statistics, which depend on the integration order of the variables, the number of regressors and the presence of deterministic components in the ARDL equation, may provide a poor approximation of the true distributions in small samples. Finite sample critical values, even if only for a sub-set of all possible model specifications, have been worked out in the literature \citep[see][]{mills2001real,narayan2004crime,kanioura2005critical,narayan2005saving}, while \cite{kripfganz2020response} provided the quantiles of the asymptotic distributions of the tests as functions of the sample size, the lag order and the number of long-run forcing variables. However, this relevant improvement does not eliminate the uncertainty related to the inconclusive regions, or the existence of other critical issues related to the underlying assumptions of the bound test framework, such as the (weak) exogeneity of the independent variables or the non-stationarity of the dependent variable.\\
As pointed out by \citet{mcnown2018bootstrapping}, some of these assumptions are often violated in empirical research. For instance, in a lot of studies the ARDL test is performed by treating each variable as dependent variable in a sequence of regressions which implicitly assume that all (or some of those) variables are endogenous. 
Other studies draw conclusions about the existence of a cointegrating relationship just by applying the overall $F$-test, without accounting for a possible occurrence of a degeneracy of first or second type (see the references quoted in \citet{mcnown2018bootstrapping} for a review of the erroneous implementation of the bound tests). \\
With the aim of addressing some drawbacks of the ARDL tests of Pesaran, Shin and Smith (PSS), such as the existence of inconclusive inference areas and distortion in size, \citet{mcnown2018bootstrapping} developed a bootstrap version of the bound tests, while \citet{sam2019augmented} derived the limiting distribution of the bound $F$-test on the lagged independent variables. 
The authors worked out the bootstrap version of the latter $F$-test, together with those of the bound $F$ overall and $t$ tests, providing an important contribution to the ARDL testing methodology. In particular the bootstrap tests, being dependent on the specific properties of the data set at hand, manage to rule out any possible inconclusive inference, which was one of the main drawback of bound tests of \citet{pesaran2001}.  Nevertheless, in their study they employed a bivariate ARDL model which does not include instantaneous differences of the independent variables in the ARDL equation, as assumed by the Pesaran approach. The absence of the latter, introduced in the ARDL equation by the operation of conditioning the dependent variable to the values of the independent ones, means that the bootstrap procedure relies on an unconditional ARDL model. Furthermore, the bivariate nature of the model does not permit to highlight the behaviour of the bootstrap tests when cointegrated independent variables appear in the error correction term.\\
In this paper, we propose a new bootstrap approach to the bound tests in a conditional ARDL model, as proposed by ~\citet{pesaran2001} in their seminal work, which goes beyond the simple bivariate model used by \citet{mcnown2018bootstrapping}.  
The importance of dealing with conditional ARDL models rests on the fact that conditioning the dependent variable, say $y_{t}$, on the others of the model, say $\bm{x}_{t}$, and assuming that $y_{t}$ does not Granger cause $\bm{x}_{t}$, makes these variables exogenous with respect to the parameters of interest of the ARDL equation. This assumption allows to rule out the VECM marginal model (i.e., the model explaining the independent variables $\bm{x}_{t}$) in the analysis of the cointegration relationship between $y_{t}$ and $\bm{x}_{t}$, since the model becomes non-informative in this respect.\\
In the paper, the performance of the bootstrap tests is evaluated via a simulation study that takes into account different data generating processes (DGPs) for the conditional ARDL model. The  DGPs cover also degenerate cases and assume the presence of either cointegrated or stationary independent variables. 
The bootstrap procedure is carried out in several steps, and it allows the construction of the (bootstrap) distribution under the null of the overall $F$ ($F_{ov}$), $t$ and $F$-test on the lagged independent variables ($F_{ind}$). \\
Inference can always be pursued with this method, also when the critical values either of the PSS tests or of the asymptotic $F$-test on the independent variables of \citet{sam2019augmented} cannot be computed.
Monte Carlo simulations confirm the better performance, in term of  size and power, of the bootstrap ARDL tests with respect to the bound ones.  By generating data under the correctly specified conditional ARDL model, and comparing the performance of the bootstrap tests estimating both the conditional and the unconditional ARDL specifications, it emerges that inference based on the latter can lead to misleading results. \\
The analysis developed in the paper focuses on two of the most frequently used specifications (restricted intercept and no trend, unrestricted intercept and no trend) of the five proposed by ~\citet{pesaran2001}.
The results herein provided open a path to an effective way of testing cointegrating relationships, with great benefit to practitioners that should avoid erroneous inference in presence of degenerate cases.\\ 
The article is organized as follows: Section \ref{sec:1sec} introduces the ARDL bound tests; Section \ref{sec:2sec} expands on the bootstrap test procedure; Section \ref{sec:3sec} delves into the Monte Carlo simulation study; Section \ref{sec:4sec} offers two illustrative applications on macroeconomic data, highlighting how inconclusive inference or an incorrect specification can be detrimental in a practical context; Section \ref{sec:5sec} concludes. 
An appendix works out the methodological aspects underlying the bound tests, offering further perspectives on some already established results.
In particular, the consequences of performing bootstrap cointegrating tests in a misspecified models are highlighted, granting a deeper understanding of the results provided by the paper.
\vspace{10mm}
\section{The ARDL Model and Bound Tests}\label{sec:1sec}
The starting point of the work of ~\citet{pesaran2001} is a $(K+1)$ VAR of order $p$
\begin{equation}\label{eq:1eq}
\bm{A}(L)(\bm{z}_t-\bm{\mu}-\bm{\eta}t)=\bm{\varepsilon}_t \enspace \enspace \enspace \bm{\varepsilon}_t\sim N(\bm{0}, \bm{\Sigma}), \enspace \enspace \enspace t=1,2,\dots,T
\end{equation}
where $\bm{A}(L)=(\bm{I}_{K+1}-  \sum_{j=1}^{p}\bm{A}_j\bm{L}^j)$. Here, $\bm{A}_i$ are square $(K+1)$ matrices, $\bm{z}_t$ a vector of  $(K+1)$ variables, $\bm{\mu}$ and $\bm{\eta}$ $(K+1)$ vectors representing drift and trend and $\bm{\varepsilon}_t$ a vector of error terms, respectively. The roots of $\bm{A}(z)$ are assumed to be either greater or equal to one.\\ To study the adjustment to equilibrium of a  variable $y_t$  given the other   $\bm{x}_t$ variables, we focus on the conditional VECM system
\begin{align}\label{eq:2eq}
\underset{(1,1)}{\Delta y_{t}}&=\alpha_{0.y}+\alpha_{1.y}t-a_{yy}{y}_{t-1}-\widetilde{\bm{a}}^{'}_{y.x}\bm{x}_{t-1}+\sum_{j=1}^{p-1}\bm{\gamma}'_{y.x,j}\Delta \bm{z}_{t-j}+\bm{\omega}'\Delta \bm{x}_{t}+\nu_{yt}\\
\label{eq:2eqb}
\underset{(K,1)}{\Delta\bm{x}_{t}}&=\bm{\alpha}_{0x}+\bm{\alpha}_{1x}t-\bm{A}_{xx}\bm{x}_{t-1}+ \sum_{j=1}^{p-1}\bm{\Gamma}_{x,j}\Delta \bm{z}_{t-j}+\bm{\varepsilon}_{xt}
\end{align}
Here, $\alpha_{0.y}$, $\alpha_{1.y}$, $\bm{\alpha}_{0x}$ and $\bm{\alpha}_{1x}$ are parameters, and $a_{yy}$, $\widetilde{\bm{a}}_{y.x}^{'}$ and $\bm{A}_{xx}$ elements of the cointegrating matrix
\begin{equation}\label{eq:4eq}
\underset{(K+1,K+1)}{\widetilde{\bm{A}}}
=\begin{bmatrix}
\underset{(1,1)}{a_{yy}} & \underset{(1,K)}{\widetilde{\bm{a}}'_{y.x}} \\ \underset{(K,1)}{\bm{0}} & \underset{(K,K)}{\bm{A}_{xx}}  \end{bmatrix},
\end{equation}
 $\sum_{j=1}^{p-1}\bm{\gamma}'_{y.x,j}\bm{L}^{j}$ and $\sum_{j=1}^{p-1}\bm{\Gamma}_{x,j}\bm{L}^{j}$ are polynomials of short-run multipliers and $\nu_{t}$ and $\bm{\varepsilon}_{xt}$ stochastic terms. \\
The ARDL equation for $\Delta y_{t}$ in \eqref{eq:2eq} can be rewritten as
\begin{equation}\label{eq:5eq}
 	\Delta y_{t}=\alpha_{0.y}+\alpha_{1.y}t -a_{yy}EC_{t-1}+ \sum_{j=1}^{p-1}\bm{\gamma}'_{j}\Delta\bm{z}_{t-j}+\bm{\omega}'\Delta\bm{x}_{t}+\nu_{yt}
\end{equation}
Here the error correction term, $EC_{t-1}$, expresses the long-run equilibrium relationship between $y_{t}$ and $\bm{x}_{t}$, while the constant $\alpha_{0.y}$ and the trend coefficient $\alpha_{1.y}$ may be not included in the equation when a restricted model is considered (see \ref{app:Appendix}).\\ 
To test the hypothesis of cointegration between $y_{t}$ and $\bm{x}_{t}$,  \citet{pesaran2001} proposed an $F$-test, $F_{ov}$ hereafter, based on the hypothesis system
\begin{align}\label{eq:7eq}
H_0: a_{yy}=0 \; \cap \;\widetilde{\bm{a}}_{y.x}=\bm{0}\\
H_1: a_{yy} \neq 0 \; \cup \;\widetilde{\bm{a}}_{y.x}\neq \bm{0}
\end{align}
Note that $H_{1}$ covers 
also the degenerate cases
\begin{align}\label{eq:7eq}
H_1^{y.x}: a_{yy}=0 \; , \;\widetilde{\bm{a}}_{y.x}\neq\bm{0}\\
H_1^{yy}: a_{yy} \neq 0 \; , \;\widetilde{\bm{a}}_{y.x} = \bm{0}
\end{align}
which are referred to as degenerate case of first and second type, respectively. Degenerate cases imply no cointegration between $y_{t}$ and $\bm{x}_{t}$.\\
The exact distribution of the $F$ statistic under the null is unknown, but it is limited from above and below by two asymptotic distributions: one corresponding to the case of stationary regressors, and another to the case of first-order integrated regressors. As a consequence, the  test is called bound test and has an inconclusive zone. Furthermore, since the test does not require all variables to be individually $I(1)$, the considered concept of long-run relationship is broader than that of cointegration. 
\footnote{The knowledge of the rank of the cointegrating matrix is necessary to overcome this impasse.}\\
 Accordingly,~\citet{pesaran2001} worked out two sets of (asymptotic) critical values: one, $\{\tau_{L,F}\}$, for the case when $\bm{x}_{t}\sim{I}(0)$ and the other, $\{\tau_{U,F}\}$, for the case when $\bm{x}_{t}\sim{I}(1)$.  These values vary in accordance with the number of regressors in  \eqref{eq:2eq}, the sample size and the assumptions made about the deterministic components (intercept and trend) of the data generating process. In this regard, five cases have been considered by the authors (see \ref{app:Appendix}). 
In case the null hypothesis is rejected, a $t$ test for the null  $a_{yy}=0$ must be carried out to rule out a degeneracy of first type. This test is bound as well, and  inconclusive inference occurs when the statistic lies between the critical values worked out for the case when all $\bm{x}_{t}$ are stationary, $\{\tau_{L,t}\}$, and first-order integrated, $\{\tau_{U,t}\}$, respectively. 
Asymptotic critical values have been worked out for this test by ~\citet{pesaran2001} for case I (no intercept and no trend), case III (unrestricted intercept, no trend) and case V (unrestricted intercept and trend), while ~\citet{nar2015} determined critical values for both the $F_{ov}$ and $t$ tests in small samples.\\
To detect the degenerate case of second type, ~\citet{mcnown2018bootstrapping} proposed to verify either the significance of $\widetilde{\bm{a}}_{y.x}$ with an $F$ test, $F_{ind}$ hereafter, or to test the order of integration of $y_{t}$. Indeed, should the lagged variables $\bm{x}_{t-1}$ be insignificant in the error correction term, then
$\Delta y_{t}$ would depend only on its lagged values in \eqref{eq:5eq} and, accordingly, $y_{t}$ would be stationary. \\
 Furthermore, ~\citet{mcnown2018bootstrapping} proposed also bootstrap $F$ and $t$ tests with the aim of evaluating the performance of the ARDL bound tests and overcoming the issue of inconclusive inference. 
By using Monte Carlo simulations,~\citet{mcnown2018bootstrapping} showed that bootstrap tests perform well in terms of both power and size, overcoming the most serious size distortions shown by the ARDL tests. Their bootstrap procedure hinges on a bivariate ARDL model, which for case III is specified as follows
\begin{equation}\label{eq:8eq}
\Delta y_{t}=\phi_{0}-a_{yy}y_{t-1}-a'_{yx}x_{t-1}+\phi_{1y}\Delta y_{t-1}+\phi_{2y}\Delta x_{t-1}+\varepsilon_{yt}
\end{equation}
\begin{equation*}
 \Delta x_{t}=\alpha_{0x}-\alpha_{xx}x_{t-1}+\phi_{1x}\Delta y_{t-1}+\phi_{2x}\Delta x_{t-1}+\varepsilon_{xt}
\end{equation*}
This is the unconditional form of the ARDL equation for $\Delta y_{t}$, as it does not include any instantaneous difference of the variable $x_{t}$, as it should be expected as consequence of the operation of conditioning $y_{t}$ on $x_{t}$. 
In addition, this simple bivariate model does not allow to evaluate the performance of the bootstrap test  when cointegrated explanatory variables are present in the EC term.\\
The model \eqref{eq:8eq} is initially estimated via OLS and the related test statistics $F_{ov}$, $t$ and $F_{ind}$ are obtained.
Subsequently, the model is re-estimated under each of the following null hypotheses 
\begin{align}\label{eq:9eq}
 F_{ov}&\rightarrow H_0:a_{yy}=0\;\cap\; a_{yx}=0\\
 t&\rightarrow H_0:a_{yy}=0\\
 F_{ind}&\rightarrow H_0:a_{yx}=0
 \end{align}
and the residuals, $\widehat{\varepsilon}_{yt}$,  $\widehat{\varepsilon}_{xt}$ are used to generate the bootstrap observations $y^{*}_{t}$, $x^{*}_{t}$ sequentially via \eqref{eq:8eq} under each null. The latter observations are in turn employed to construct the  bootstrap distributions $F_{ov}^{H_0}$, $t^{H_0}$ and $F_{ind}^{H_0}$, by repeating the procedure a large number of times. Finally, the (bootstrap) $p$-values related to the $F_{ov}$, $t$ and $F_{ind}$ statistics previously obtained are evaluated by using the bootstrap distributions under their respective null.\\ 
In Section \ref{sec:2sec} we propose a bootstrap alternative to the bound tests that relies on the conditional version of the ARDL equation for $\Delta y_t$, coherently with the approach proposed by ~\citet{pesaran2001}. 
Monte Carlo simulations are then carried out to investigate the performance of our proposal which, assuming a multivariate DGP for the ARDL model, hinges on a more complex simulation context than the one assumed by~\citet{mcnown2018bootstrapping}.
\vspace{10mm}
\section{The New Bootstrap Procedure}\label{sec:2sec}
The bootstrap procedure here proposed focuses on ARDL models with no trend and either restricted or unrestricted intercept, referred to as case II and case III respectively (see \ref{app:Appendix}).
\begin{itemize}
\item Case II 
\begin{equation}\label{eq: 3.1a}
\Delta y_{t}=-a_{yy}(y_{t-1}-\mu_{y})-\widetilde{\bm{a}}^{'}_{y.x}(\bm{x}_{t-1}-\bm{\mu}_{x})+ \sum_{j=1}^{p-1}\bm{\gamma}_{y.x,j}'\Delta\bm{z}_{t-j}+ \bm{\omega}'\Delta\bm{x}_{t}+\nu_{yt}
\end{equation}
\item Case III 
\begin{equation}\label{eq: 3.11b}
\Delta y_{t}=\alpha_{0.y}-a_{yy} y_{t-1}-\widetilde{\bm{a}}^{'}_{y.x}\bm{x}_{t-1}+ \sum_{j=1}^{p-1}\bm{\gamma}_{y.x,j}'\Delta\bm{z}_{t-j}+ \bm{\omega}'\Delta\bm{x}_{t}+\nu_{yt}
\end{equation}
\end{itemize}
Two independent variables have been included in the model and different data generating processes (DGPs) have been considered in order to duly take into account a variety of situations implying cointegration, no cointegration and degenerate cases.\\
The bootstrap procedure consists of the following steps:\\
\begin{enumerate}
\item The unrestricted ARDL model, \eqref{eq: 3.1a} or \eqref{eq: 3.11b}, is estimated via OLS and the related test statistics $F_{ov}$, $t$ or $F_{ind}$ are computed. For case II, the intercept is also restricted to be zero under the null of the $F_{ov}$ test, differently from case III where the intercept does not partake the test (see \ref{app:Appendix}). 
\item In order to construct the distribution of each test statistic under the respective null, the same model is re-estimated imposing the appropriate restrictions on the coefficients affected by each of the tests under consideration. For example, considering $F_{ind}$ and case III, it amounts to estimating the model in \eqref{eq: 3.1a} disregarding $\bm{x}_{t-1}$ from the set of explanatory variables of $\Delta y_{t}$.
\item The restricted residuals are computed. For example, regarding case III, the residuals are
\begin{equation}\label{eq:14}
\widehat{\nu}_{yt}^{F_{ov}}=\Delta y_{t}-\sum_{j=1}^{p-1}\widehat{\bm{\gamma}}_{y.x,j}'\Delta\bm{z}_{t-j}-\widehat{\bm{\omega}}'\Delta\bm{x}_{t}
\end{equation}
\begin{equation}\label{eq:15}
\widehat{\nu}_{yt}^{t}=\Delta y_{t}+\widehat{\widetilde{\bm{a}}}'_{y.x}(\bm{x}_{t-1}-\widehat{\bm{\mu}}_{x})-  \sum_{j=1}^{p-1}\widehat{\bm{\gamma}}_{y.x,j}'\Delta\bm{z}_{t-j}-\widehat{\bm{\omega}}'\Delta\bm{x}_{t}
\end{equation}
\begin{equation}\label{eq:16}
\widehat{\nu}_{yt}^{F_{ind}}=\Delta y_{t}+\widehat{a}_{yy}(y_{t-1}-\widehat{\mu}_{y})-  \sum_{j=1}^{p-1}\widehat{\bm{\gamma}}_{y.x,j}'\Delta\bm{z}_{t-j}-\widehat{\bm{\omega}}'\Delta\bm{x}_{t}
\end{equation}
Here, the apex "$\hat{.}$" denotes the estimated parameters. Case II can be dealt with analogously.
\item The marginal VECM model explaining the independent variables 
\begin{equation}\label{eq:17eq}
 \Delta\bm{z}_{t}=\bm{\alpha}_{0}-\bm{A}\bm{z}_{t-1}+ \sum_{j=1}^{p-1}\bm{\Gamma}_{j}\Delta\bm{z}_{t-j}+\bm{\varepsilon}_{t}
\end{equation}
is estimated, and the residuals  
\begin{equation}\label{eq:18eq}
\widehat{\bm{\varepsilon}}_{xt}= \Delta\bm{x}_{t}-\widehat{\bm{\alpha}}_{0x}+\widehat{\bm{A}}_{xx}\bm{x}_{t-1}- \sum_{j=1}^{p-1}\widehat{\bm{\Gamma}}_{(x)j}\Delta\bm{z}_{t-j}.
\end{equation}
 are thus computed. This approach guarantees that $\widehat{\bm{\varepsilon}}_{xt}$ are uncorrelated with the ARDL residuals $\widehat{\nu}_{yt}^{.}$.
 \item A large set of $B$ bootstrap replicates are extracted from the residuals calculated as in \eqref{eq:14},\eqref{eq:15}, \eqref{eq:16} and \eqref{eq:18eq}. In each replication, the following operations are carried out

\begin{enumerate}
 \item Each set of $(T-p)$ resampled residuals (with replacement) $\widehat{\bm{\nu}}_{zt}^{(b)}=(\widehat{\nu}_{yt}^{(b)},\widehat{\bm{\varepsilon}}_{xt}^{(b)})$ is re-centered \citep[see][]{davidson2005case}
\begin{align}\label{eq:19eq}
\dot{\widehat{\nu}}^{(b)}_{yt}&=\widehat{\nu}^{(b)}_{yt} -\frac{1}{T-p}\sum_{t=p+1}^{T}\widehat{\nu}^{(b)}_{yt}\\\label{eq:20eq}
\dot{\widehat{\bm{\varepsilon}}}^{b}_{x_{i}t}&=\widehat{\bm{\varepsilon}}^{(b)}_{x_{i}t}-\frac{1}{T-p}\sum_{t=p+1}^{T}\widehat{\bm{\varepsilon}}^{(b)}_{x_{i}t}\qquad i=1,\dots,K
\end{align}
\item A sequential set of $(T-p)$ bootstrap observations, $y^{*}_{t}\enspace, \bm{x}^{*}_{t}\enspace t=p+1,\dots,T$, is generated as follows
 \begin{equation}
 y^{*}_{t}=y^{*}_{t-1}+\Delta y^{*}_{t}, \enspace \enspace \bm{x}^{*}_{t}=\bm{x}^{*}_{t-1}+\Delta \bm{x}^{*}_{t}
\end{equation}
where $\Delta \bm{x}^{*}_{t}$ are obtained from \eqref{eq:18eq} and  $\Delta y^{*}_{t}$  from either \eqref{eq:14}, \eqref{eq:15} or \eqref{eq:16} after replacing in each of these equations the original residuals with the bootstrap ones. \\
The initial conditions, that is the observations before $t=p+1$ are obtained by drawing randomly $p$ observations in block from the original data, so as to preserve the data dependence structure.

\item An unrestricted ARDL model is estimated via OLS using the bootstrap observations, and the statistics $F_{ov}^{(b),H_0}$, $t^{(b),H_0}$ $F_{ind}^{(b),H_0}$ are computed.
\end{enumerate}
\item The bootstrap distributions of $\big\{F_{ov}^{(b),H_0}\big\}_{b=1}^B$, $\big\{F_{ind}^{(b),H_0}\big\}_{b=1}^B$ and $\big\{t^{(b),H_0}\big\}_{b=1}^B$ under the null are then employed to determine the critical values of the tests. By denoting with $T^*_b$ the ordered bootstrap test statistic, and with $\alpha$ the nominal significance level, the bootstrap critical values are determined as follows 
 \begin{equation}\label{eq:21eq}
c^*_{1-\alpha}=\min\bigg\{c:\sum_{b=1}^{B}\bm{1}_{\{T^*_b >c\}}	\leq\alpha\bigg\}\end{equation}
for the F tests and 
\begin{equation}\label{eq:22eq}
c^*_{{\alpha}}=\max\bigg\{c:\sum_{b=1}^{B}\bm{1}_{\{T^*_b<c\}}	\leq {\alpha}\bigg\}\end{equation}
for the $t$ test.\\
Here, $\bm{1}_{\{x \in A\}}$ is the indicator function, which is equal to one if the condition in subscript is satisfied and zero otherwise. 
\end{enumerate}
The null hypothesis is rejected if the F statistic computed at step 1, $F_{ov}$ or $F_{ind}$, is greater than $c^*_{1-\alpha}$, or if the $t$ statistic computed at the same step is lower than $c^*_{{\alpha}}$.
\vspace{10mm}
\section{Results from Monte Carlo Simulations}\label{sec:3sec}
In this section, a simulation study on three variables, $(y_{t}, x_{1t}, x_{2t})$, has been carried out with the aim of evaluating the performance of the bootstrap test in detecting cointegration between $y_{t}$ and the other variables, $\bm{x}_{t}=(x_{1,t}, x_{2,t})$, in an ARDL model. 
The data for these variables have been generated starting from the trivariate density of the error term $\boldsymbol \varepsilon_t=(\varepsilon_{yt},\varepsilon_{x_1t},\varepsilon_{x_2t})$ of the VAR model in Equation \eqref{eq:1eq}. Hence, using the values of the variance/covariance matrix, $\bm{\Sigma}$, of $\boldsymbol \varepsilon_t$, the conditional error term of the ARDL equation,  $\nu_{yt}=\varepsilon_{yt}-\boldsymbol\omega'\boldsymbol\varepsilon_{xt}$, has been computed (see \ref{app:Appendix}). Afterwards, using the conditional VECM system of Equations \eqref{eq:2eq}-\eqref{eq:2eqb}, the first differences of the explanatory variables, $\Delta \bm{x}_{t}$, have been generated, and the latter used to obtain $\Delta {y}_{t}$. In order to preserve temporal dependence, the data have been generated recursively, one observation at a time. A burn-in period of 50 observations has been considered, using the first $(p-1)$ rows of $(\nu_{yt},\boldsymbol\varepsilon_{xt})$ as starting values.\\
\color{black}
Different DGPs for the conditional ARDL models have been considered for cases II (restricted intercept and no trend) and III (unrestricted intercept and no trend), specified in 
\eqref{eq: 3.1a} and \eqref{eq: 3.11b}. 
 For comparison reasons, the tests, $F_{ov}$, $F_{ind}$, $t$, in misspecified models, the tests have been carried out in both the unconditional and conditional versions of the ARDL model for cases II and III.
 This means that, by taking for instance case III given in Equation \eqref{eq: 3.11b}, the tests have been performed on the parameters of both the following  specifications
\begin{align}
\label{eq: 3.1b}
\text{(C)}&\qquad \Delta y_{t}=\alpha_{0.y}-a_{yy} y_{t-1}-\widetilde{\bm{a}}^{'}_{y.x}\bm{x}_{t-1}+ \sum_{j=1}^{p-1}\bm{\gamma}_{y.x,j}'\Delta\bm{z}_{t-j}+ \bm{\omega}'\Delta\bm{x}_{t}+\nu_{yt}\\
\label{eq: 3.1b}
\text{(UC)}& \qquad\Delta y_{t}= \alpha_{0y} -a_{yy} y_{t-1}-\bm{a}^{'}_{yx}\bm{x}_{t-1}+ \sum_{j=1}^{p-1}\bm{\gamma}_{y,j}'\Delta\bm{z}_{t-j}+
\varepsilon_{yt}
\end{align} 
Here,
\begin{equation}
\widetilde{\bm{a}}^{'}_{y.x}=\bm{a}^{'}_{yx}-\boldsymbol\omega'\bm{A}_{xx}=\bm{a}^{(UC)'}_{yx}-\bm{a}^{(C)'}_{yx}
\end{equation}
with $\bm{a}^{(UC)'}_{yx}=\bm{a}^{'}_{yx}$ and $\bm{a}^{(C)'}_{yx}=\boldsymbol\omega'\bm{A}_{xx}$. It is worth noting that the term $\bm{a}^{(C)'}_{yx}=\bm{\omega}^{'}\bm{A}_{xx}$ does not appear in the estimation of the unconditional model, because it is the term which is introduced in the conditional ARDL, by conditioning $y_{t}$ on the independent variables $\bm{x}_{t}$. \\
The error covariance matrix $\boldsymbol\Sigma$ of the VAR model \eqref{eq:1eq}, and the parameter vector $\boldsymbol\omega'=\boldsymbol\sigma_{yx}'\boldsymbol{\Sigma}_{xx}^{-1}$ resulting from conditioning have been specified as follows
\begin{equation*}
\boldsymbol\Sigma =
\begin{bsmallmatrix}
1.69&&\\
0.39&1.44&\\
0.52&-0.3&1
\end{bsmallmatrix}\qquad
\boldsymbol\omega' = \left[ 0.40\overline 4\;\;0.641\overline 3\right]
\end{equation*}
Regarding the short-run part of the unconditional VECM and conditional ARDL equations, $p=2$ lag periods have been chosen, with unconditional parameter matrices 
$\boldsymbol\Gamma_1$ and $\boldsymbol\Gamma_2$ and conditional parameter vectors $\bm{\gamma}^{'}_{y.x,j}=\bm{\gamma}_{y,j}'-\bm{\omega}'\bm{\Gamma}_{(x),j}$, $j=1,2$, specified as follows\color{black}
\begin{align*}
\boldsymbol\Gamma_1 &=
\begin{bsmallmatrix}
0.6&0&0.2\\
0.1&-0.3&0\\
0&-0.3&0.2
\end{bsmallmatrix}\qquad\qquad\qquad\qquad\;\;\,
\boldsymbol\Gamma_2 =
\begin{bsmallmatrix}
0.2&0&0.1\\
0.05&-0.15&0\\
0&0&0.1
\end{bsmallmatrix}\\
\boldsymbol\gamma_{y.x,1}' &= \left[0.559\overline 5\;\;0.317\overline 3\;\;0.0717\overline 3\right]\qquad
\boldsymbol\gamma_{y.x,2}' = \left[ 0.179\overline 7\;\;0.060\overline 3\;\;0.0358\overline 6\right].
\end{align*}
The conditional VECM intercept is $\boldsymbol\alpha_0^c = \widetilde{\bm A}\boldsymbol\mu$. Here, $\boldsymbol\mu'=(\mu_{y}, \mu_{x_{1}},\mu_{x_{2}} )=[0.2\;0.3\;0.4]$ is the mean vector of the VAR model \eqref{eq:1eq}, and $\widetilde{\bm A}$ is the cointegrating matrix resulting from the operation of conditioning $y_{t}$ on $\bm{x}_{t}$ and setting $\bm{a}_{xy}=\bm{0}$\\
  \begin{equation}\label{simA}
\widetilde{\bm{A}}=\begin{bmatrix}a_{yy}, & \widetilde{\bm{a}}_{y.x}\\
\bm{0},& \bm{A}_{xx} \end{bmatrix}.
\end{equation}\\
Under case II, the conditional ARDL intercept is $\alpha_{0.y}=\bm e_1'\boldsymbol\alpha_0^c$, with $\bm e_1=[1\;0\;0]$, and is tied to the long-run relationships between the (possibly) cointegrated variables. On the contrary, under case III the intercept is unrestricted, and set to $\alpha_{0.y}=0.3$.\color{black}\\
The following specifications have been assumed for the sub-matrix $\boldsymbol A_{xx}$ in \eqref{simA}
\begin{equation}
\begin{array}{ll}
\text{case A)} \enspace \boldsymbol A_{xx} = \begin{bsmallmatrix}
0\\
0.7
\end{bsmallmatrix}\begin{bsmallmatrix}
1.1&
1.1
\end{bsmallmatrix}\\
\text{case B)} \enspace \boldsymbol A_{xx} = \begin{bsmallmatrix}
0.3&-0.4\\
0.5&0.3
\end{bsmallmatrix}\\
\end{array}
\end{equation}
depending on whether the independent variables have been assumed to be cointegrated (A), or not cointegrated, namely stationary (B).\\
As for the elements $a_{yy}$, $\bm a^{(UC)'}_{yx}$, $\bm a^{(C)'}_{yx}$, they depend on the DGP under consideration. When they are not assumed to be null, they are set equal to
\begin{equation}\label{eq:parspec}
\begin{array}{ll}
\qquad\;\;\; a_{yy}=0.7\qquad \bm a^{(UC)'}_{yx}=[0.6, \; 0.4]\vspace{0.3cm}\\
\text{A)} \enspace \bm{a}^{(C)'}_{yx}=\begin{bsmallmatrix}
0.4938&0.4938 
\end{bsmallmatrix} \qquad \bm{a}_{y.x}'=\begin{bsmallmatrix}
0.1062&-0.0938 
\end{bsmallmatrix}\\
\text{B)} \enspace \bm{a}^{(C)'}_{yx}= \begin{bsmallmatrix}
0.442&0.0306\overline 2
\end{bsmallmatrix}\qquad \bm{a}_{y.x}'=\begin{bsmallmatrix}
0.158&0.3693\overline 7
\end{bsmallmatrix}\\
\end{array}
\end{equation}
Accordingly, the DGPs under examination are
\begin{itemize}
\item DGP 1: 
$a_{yy} \neq 0$\enspace,\enspace $\bm a^{(UC)'}_{yx}\neq \bm 0$, \enspace\enspace $\rightarrow$ \enspace $\widetilde{\bm{a}}^{'}_{y.x}=\bm a^{(UC)'}_{yx}- \bm{a}^{(C)'}_{yx}\neq \bm 0$.
\item DGP 2: 
$a_{yy} = 0$,\enspace \enspace$\bm a^{(UC)'}_{yx}\neq \bm 0$,\enspace \enspace $\rightarrow$ \enspace $\widetilde{\bm{a}}^{'}_{y.x}=\bm a^{(UC)'}_{yx}- \bm{a}^{(C)'}_{yx}\neq \bm 0$.
\item DGP 3: 
$a_{yy} = 0$, \enspace \enspace $\bm a^{(UC)'}_{yx}=\bm{a}^{(C)'}_{yx}$ \enspace $\rightarrow$ \enspace $\widetilde{\bm{a}}^{'}_{y.x} = \bm 0$.
\item DGP 4:
$a_{yy} = 0$,\enspace \enspace$\bm a^{(UC)'}_{yx}=\bm 0$ \enspace $\rightarrow$ \enspace $\widetilde{\bm{a}}^{'}_{y.x}=\bm{a}^{(C)'}_{yx}\neq \bm 0$.
\item DGP 5: 
$a_{yy} \neq 0$,\enspace \enspace$\bm a^{(UC)'}_{yx}=\bm 0$\enspace $\rightarrow$ \enspace $\widetilde{\bm{a}}^{'}_{y.x}=\bm{a}^{(C)'}_{yx}\neq \bm 0$.
\item DGP 6:
$a_{yy} \neq 0$,\enspace \enspace$\bm a^{(UC)'}_{yx}=\bm{a}^{(C)'}_{yx}$\enspace $\rightarrow$ \enspace $\widetilde{\bm{a}}^{'}_{y.x} = \bm 0$.
\end{itemize}

DGP 1 implies the existence of cointegration between $y_{t}$ and the $\bm{x}_{t}$ variables. The other DGPs deal with degenerate cases (see \ref{app:Appendix}).
To account for different levels of strength in cointegration, two configurations are assumed for this DGP: the former, denoted DGP 1H, is given in  Equation \eqref{eq:parspec}, the latter, denoted DGP 1L, is as in  \eqref{eq:parspec} but with values of $a_{yy}$ and $\bm{a}_{yx}^{(UC)}$ which are halved with respect to the ones of the former. \\ Under DGPs 2, 3 and 4, both the cases of cointegrated independent variables (case A, $rk(\bm{A}_{xx})=1$) and stationary independent variables (case B, $rk(\bm{A}_{xx})=2$) are considered. \\
In particular, under DGP 3 degeneracy cases of both first and second type occur ($a_{yy}=0, \; \widetilde{\bm{a}}^{'}_{y.x}=\bm 0)$. It is worth noting that an analysis based on the unconditional ARDL does not allow to detect the degeneracy of second type as in this model $\widetilde{\bm{a}}^{'}_{y.x}=\bm{a}^{(UC)'}_{yx}\neq \bm 0$.\\
Under DGP 4, only a degeneracy of first type occurs ($a_{yy}=0$). Similarly, in this setting an analysis based on the unconditional model would lead to the incorrect conclusion that degenerate cases of both first and second type occur ($a_{yy}=0, \; \widetilde{\bm{a}}^{'}_{y.x}=\bm{a}^{(UC)'}_{yx}=\bm 0)$. \\
Regarding DGP 5, no degeneracy case is at work ($a_{yy} \neq 0, \; \widetilde{\bm{a}}^{'}_{y.x}\neq\bm 0$). However, as the only term in the unlagged $\bm{x}_{t}$ appearing in the ARDL equation is the stationary term introduced by conditioning $y_{t}$ on $\bm{x}_{t}$, namely  $\bm{a}^{(C)'}_{yx}\bm{x}_{t-1}=\bm{\omega}'\bm{A}_{xx}\bm{x}_{t-1}$, $y_{t}$ must be stationary as well (see \ref{app:Appendix}). It is worth noting that according to the unconditional model a degeneracy case of second type is at work ($a_{yy} \neq 0, \; \widetilde{\bm{a}}^{'}_{y.x}=\bm{a}^{(UC)'}_{yx}=\bm 0)$.\\
Finally, under DGP 6 a degeneracy case of second type occurs ($a_{yy} \neq 0, \; \widetilde{\bm{a}}^{'}_{y.x}=\bm 0$), since the term $\bm{a}^{(C)'}_{yx}$, introduced in the ARDL equation by conditioning $y_{t}$ on $\bm{x}_{t}$, cancels off $\bm{a}^{(UC)'}_{yx}$. On the contrary, in the unconditional model no degeneracy case occurs as $a_{yy} \neq 0, \; \widetilde{\bm{a}}^{'}_{y.x}=\bm{a}^{(UC)'}_{yx} \neq\bm 0$ \\ 
It is worth noting that under DGPs 1, 5 and 6, the case of stationary independent variables (case B) is not considered. Indeed, upon noting that when $a_{yy} \neq 0$
\begin{equation}
rk(\bm{A})=1+rk(\bm{A}_{xx})
\end{equation}
it becomes clear that it is possible to consider the case of stationary variables only when a degeneracy of first type occurs. Otherwise, the matrix $\bm{A}$ would be of full rank, thus ruling out the existence of cointegrating relationships among the remaining independent variables.\\
Table \ref{tab:sim1} provides the results of both the ARDL and bootstrap tests when performed in the conditional and unconditional ARDL equation for cases II and  III,  with a sample size of $T=200$, and $K=1000$ replications.
\\
In the table, values not in brackets represent the frequency of rejection of the null hypothesis. For this reason, when the assumption of the DGP is coherent with the null, we use the term size / coverage. Conversely, when the assumption of the DGP is not coherent with the null, we use the term power. As for the null hypothesis, it is formulated on the basis of the parametrization of the conditional model, which is assumed to be the correct one.
\begin{itemize}
    \item \textit{\textbf{DGP 1}: $H_0$ false for all tests.}\\
    Both bound and bootstrap tests, $F_{ov}$ and $t$, perform well. As expected, the power of the tests is high in both the conditional and unconditional ARDL equation. The same occurs for the bootstrap $F_{ind}$ test and the asymptotic $F$ test on the independent variables proposed by \citet{sam2019augmented} for case III, denoted by $F_{ind}^{SMG}$.The power of the $F_{ind}$ test slightly decreases for DGP 1L, and this decrement is more evident in the UC model.
    \item \textit{\textbf{DGP 2}:  $H_0$ true for $t$, false for $F_{ov}$ and $F_{ind}$.}
    The $F_{ov}$ test, either bootstrap or bound, mostly rejects the null in both the conditional and unconditional ARDL equation. The bootstrap $t$ test signals the nullity of $a_{yy}$ with a coverage that, at the significance level $\alpha=0.05$, is much better than that of the bound $t$ test, available only for case III.
    \item \textit{\textbf{DGP 3}: $H_0$ true for all tests.}
    Under this case, degeneracy of first and second type are at work simultaneously.  
    The $F_{ov}$ test, either bootstrap or bound, rejects the null with a coverage that is much better in the conditional rather than the unconditional model. Furthermore, the performance of the bootstrap  $F_{ov}$ is better in comparison with that of the corresponding bound version. 
    The coverage of the bootstrap $t$ test is closer to the nominal one than that of the bound test in case III. The $F_{ind}$ test correctly signals the absence of the explanatory variables in lagged levels in the conditional model, while it over-rejects the null in the unconditional one. In  case III, the test in the conditional model has a better coverage than the corresponding $F_{ind}^{SMG}$.
    \item \textit{\textbf{DGP 4}: $H_0$ true for $t$, false for $F_{ov}$ and $F_{ind}$.}\\
     The $F_{ov}$ test, either bootstrap or bound, correctly rejects the null in the conditional ARDL model. The  power is higher for the bootstrap test than for the bound one and, particular, it is higher in the conditional model rather than in the unconditional one. The bootstrap $t$ test shows a better coverage than the bound one for case III. The bootstrap $F_{ind}$ test rejects the null in the conditional ARDL with a power which is higher than that of the corresponding test in the unconditional model. A similar behavior is shown by the $F_{ind}^{SMG}$ test for case III, albeit with lower power than the bootstrap one.
    \item \textit{\textbf{DGP 5}: $H_0$ false for all tests.}\\
    The $F_{ov}$ and $t$ tests, either bound or bootstrap, strongly reject the null in  both the conditional and unconditional models. The $F_{ind}$ test leads to contrasting results in the conditional and unconditional models. The power of the test in the former model is much higher than that in the latter and in each of these models the performance of the bootstrap $F_{ind}$ test  is better than that of the $F_{ind}^{SMG}$ test.
    \item \textit{\textbf{DGP 6}: $H_0$ false for $F_{ov}$ and $t$, true for $F_{ind}$.}\\
    The $F_{ov}$ test, in both the bound and bootstrap versions, strongly rejects the null. The same happens for the $t$ test, either bound or bootstrap. The $F_{ind}$ incorrectly rejects the null with a coverage (size) that is too high in the unconditional ARDL, while it has a correct size when performed in  the conditional ARDL. A similar performance is shown by the  $F_{ind}^{SMG}$ for case III, albeit with more inaccuracy.
    \end{itemize}
    For DGPs 2, 3 and 4, both the cases of co-integrated ($rk(\bm{A}_{xx})<k$) and stationary ($rk(\bm{A}_{xx})=k$) independent variables have been considered. In general, no great differences have been registered between the test outcomes for these two cases, except that either the size or the power of the bootstrap and bound tests is slightly better when the explanatory variables are cointegrated.\\
\color{black}

\begin{figure}[hp!]
    \centering
    \includegraphics[scale=0.7]{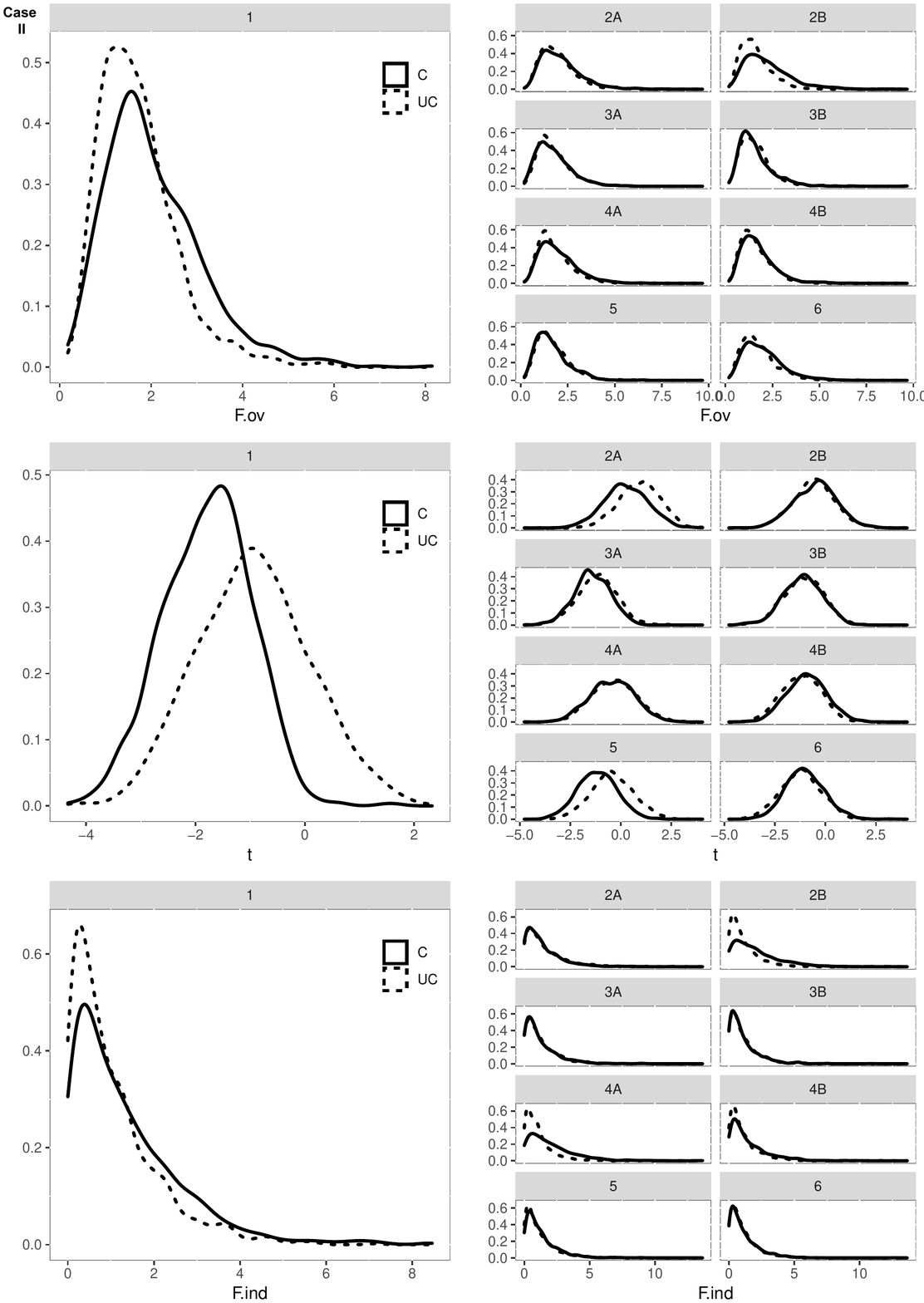}
     \caption{Bootstrap distributions of the statistics $F_{ov}^{H_0}$, $t^{H_0}$ and $F_{ind}^{H_0}$, case II. Panel titles refer to the DGP, line type refers to either the conditional (C) or unconditional (UC) specification.}
     \label{fig:boot1}
\end{figure}

\begin{figure}[hp!]
    \centering
    \includegraphics[scale=0.7]{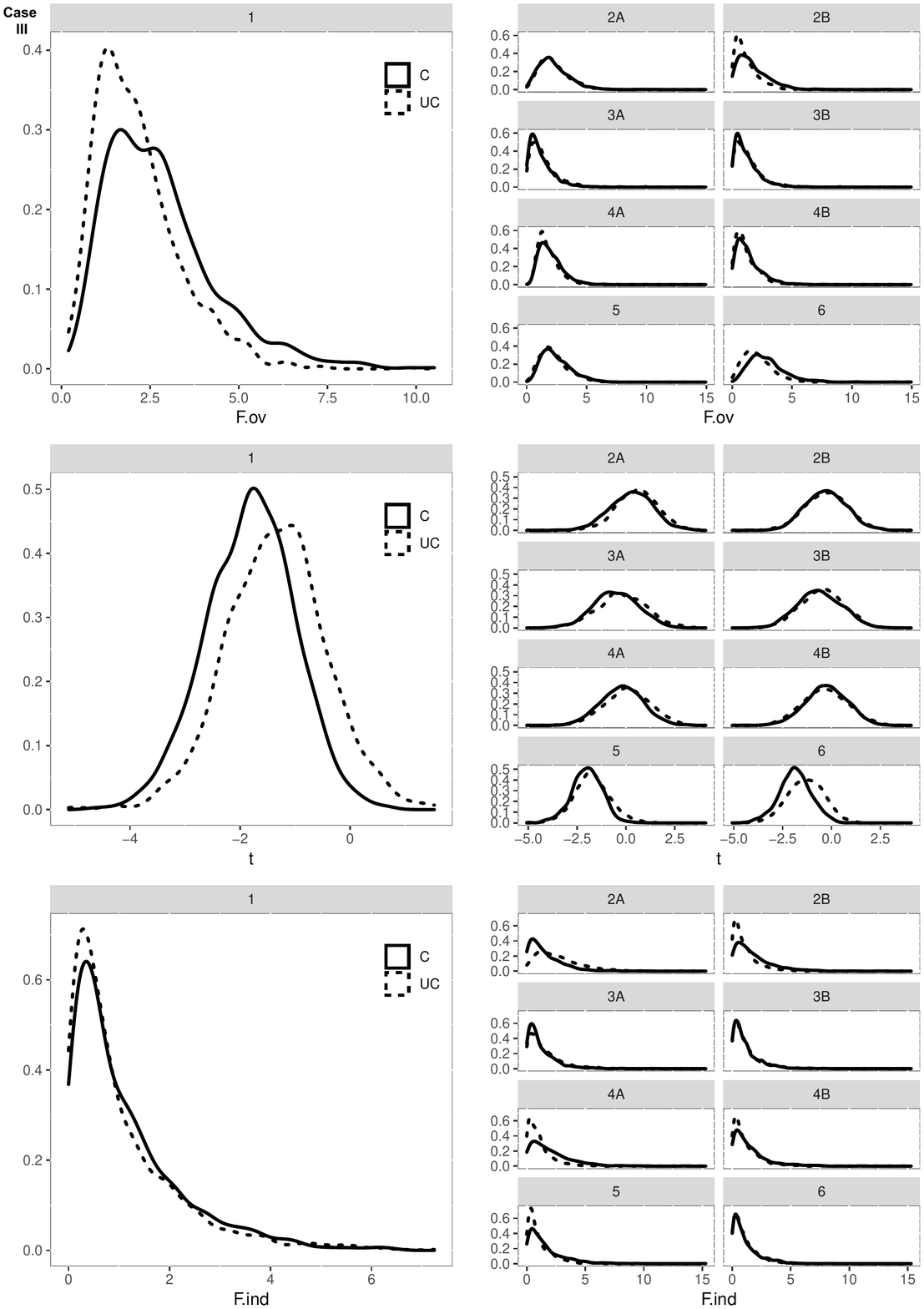}
    \caption{Distributions of the bootstrap statistics $F_{ov}^{H_0}$, $t^{H_0}$ and $F_{ind}^{H_0}$, case III. Panel titles refer to the DGP, line type refers to either the conditional (C) or unconditional (UC) specification.}
    \label{fig:boot2}
\end{figure}

\begin{table}
  \resizebox{\textwidth}{!}{
    \begin{tabular}{cccccccllllll}
    Case&Spec & {DGP} &$rk(\bm A_{xx})$& $a_{yy}$   & $\bm{a}_{yx}^{(UC)} $ & $\widetilde{\bm a}_{y.x}$   
    & $F_{ov}$ (PSS) & $F_{ov}$ (Boot) & $t$ (PSS)& $t$ (Boot)& $F_{ind}$ (SMG) & $F_{ind}$ (Boot) \\\midrule
    \multirow{20}[3]{*}{Case II} & \multirow{10}[1]{*}{Conditional} 
    
    &{1H} & < K   & $ \neq 0$    & $ \neq\bm 0$ & $ \neq\bm 0$ & 0.979* (0.014) & 1* & -     & 1* & -     & 0.997* \\
          &       & {1L} & {<K} & $ \neq 0$    & $ \neq\bm 0$ & $ \neq\bm 0$& 1* (0) & 1* & -     & 1* & -     & 0.908* \\
          &       & {2A} & {<K} & {0} & $ \neq\bm 0$ & $ \neq\bm 0$ & 0.982* (0.015) & 0.984* & -     & 0.051 & -     & 0.990* \\
          &       & 2B    & = K   & {0} & $ \neq\bm 0$ & $ \neq\bm 0$ & 1* (0) & 0.989* & -     & 0.060 & -     & 0.993* \\
          &       & {3A} & {< K} & {0} & $ \neq\bm 0$ & $ =\bm 0$ & 0.033 (0.064) & 0.043 & -     & 0.054 & -     & 0.088 \\
          &       & {3B} & = K   &{0} & $ \neq\bm 0$ & $ =\bm 0$ & 0.018 (0.031) & 0.036 & -     & 0.038 & -     & 0.030 \\
          &       & {4A} & {< K} & {0} & $ =\bm 0$ & $ \neq\bm 0$ & 0.768* (0.12) & 0.740* & -     & 0.044 & -     & 0.848* \\
          &       & {4B} & = K   &{0} & $ =\bm 0$ & $ \neq\bm 0$ & 0.552* (0.178) & 0.657* & -     & 0.043 & -     & 0.796* \\
          &       & 5     & { < K} & $ \neq 0$    & $ =\bm 0$ & $ \neq\bm 0$ & 1* (0) & 1* & -     & 1* & -     & 0.901* \\
          &       & 6     & { < K} & $ \neq 0$    & $ \neq\bm 0$ & $ =\bm 0$ & 1* (0) & 1* & -     & 1* & -     & 0.046 \\
\cmidrule{2-13}          & \multirow{10}[2]{*}{Unconditional} 
&{1H} & < K   & $ \neq 0$    & $ \neq\bm 0$ & $ \neq\bm 0$ & 1* (0) & 1* & -     & 1* & -     & 1* \\
        &       & {1L} & {<K} & $ \neq 0$    & $ \neq\bm 0$ & $ \neq\bm 0$& 1* (0)  & 1* & -    & 1* & -     &0.896*  \\
        &       & {2A} & {< K} & {0} & $ \neq\bm 0$ & $ \neq\bm 0$ & 1* (0) & 1* & -     & 0.051 & -     & 1* \\
          &       & 2B    & = K   & {0} & $ \neq\bm 0$ & $ \neq\bm 0$ & 1* (0) & 1* & -     & 0.058 & -     & 1* \\
          
          &       &{3A} & {< K} & {0} & $ \neq\bm 0$ & $ =\bm 0$ & 0.834 (0.08) & 0.879 & -     & 0.049 & -     & 0.923 \\
          &       & {3B} & = K   & {0} & $ \neq\bm 0$ & $ =\bm 0$ & 0.655 (0.143) & 0.790 & -     & 0.038 & -     & 0.903 \\
          &       & {4A} & {< K} & {0} & $ =\bm 0$ & $ \neq\bm 0$ & 0.03* (0.065) & 0.053* & -     & 0.054 & -     & 0.106* \\
          &       & {4B} & = K   & {0} & $ =\bm 0$ & $ \neq\bm 0$ & 0.011* (0.045) & 0.047* & -     & 0.036 & -     & 0.045* \\
          &       & 5     & < K   & $ \neq 0$    & $ =\bm 0$ & $ \neq\bm 0$ & 1* (0) & 1* & -     & 1* & -     & 0.063* \\
          &       & 6     & < K   & $ \neq 0$    & $ \neq\bm 0$ & $ =\bm 0$ & 1* (0) & 1* & -     & 1* & -     & 0.964 \\
    \midrule
    \multirow{20}[4]{*}{Case III} & \multirow{10}[2]{*}{Conditional} 
    & {1H} & < K   & $ \neq 0$    & $ \neq\bm 0$ & $ \neq\bm 0$ & 1* (0) & 1* & 1* (0) & 1* & 0.992* (0.006) & 0.998* \\
     &       & {1L} & {<K} & $ \neq 0$    & $ \neq\bm 0$ & $ \neq\bm 0$& 1* (0) & 1* &  1* (0)    & 1*  &  0.742* (0.206)   & 0.906*\\
          &       & {2A} & {< K} & {0} & $ \neq\bm 0$ & $ \neq\bm 0$ & 0.979* (0.003) & 0.998* & 0 (0) & 0.060 & 0.972* (0.026) & 0.991* \\
          &       & 2B    & = K   & {0} & $ \neq\bm 0$ & $ \neq\bm 0$ & 0.971* (0.011) & 0.986* & 0.023 (0) & 0.041 & 0.973* (0.025) & 0.994* \\
          &       & {3A} & {< K} & {0} & $ \neq\bm 0$ & $ =\bm 0$ & 0.014 (0.018) & 0.043 & 0.019 (0) & 0.060 & 0.016 (0.089) & 0.051 \\
          &       & {3B} & = K   & {0} & $ \neq\bm 0$ & $ =\bm 0$ & 0.004 (0.017) & 0.042 & 0.016 (0) & 0.060 & 0.006 (0.049) & 0.026 \\
          &       & {4A} & {< K} & {0} & $ =\bm 0$ & $ \neq\bm 0$ & 0.739* (0.122) & 0.911* & 0.009 (0) & 0.071 & 0.833* (0.128) & 0.935* \\
          &       & {4B} & = K   & {0} & $ =\bm 0$ & $ \neq\bm 0$ & 0.492* (0.178) & 0.731* & 0.008 (0) & 0.056 & 0.638* (0.245) & 0.806* \\
          &       & 5     & < K   & $ \neq 0$    & $ =\bm 0$ & $ \neq\bm 0$ & 1* (0) & 1* & 1* (0) & 1* & 0.734* (0.197) & 0.885* \\
          &       & 6     & < K   & $ \neq 0$    & $ \neq\bm 0$ & $ =\bm 0$ & 1* (0) & 1* & 1* (0) & 1* & 0.004 (0.046) & 0.035 \\
\cmidrule{2-13}          & \multirow{10}[2]{*}{Unconditional} 
& {1H} & < K   & $ \neq 0$    & $ \neq\bm 0$ & $ \neq\bm 0$ & 1* (0) & 1* & 1* (0) & 1* & 0.998* (0) & 1* \\
          &       & {1L} & {<K} & $ \neq 0$    & $ \neq\bm 0$ & $ \neq\bm 0$& 1* (0) & 1*& 1* (0)   &1*  & 0.708* (0.222)     & 0.890*\\
          &       & {2A} & {< K} & {0} & $ \neq\bm 0$ & $ \neq\bm 0$ & 0.997* (0.003) & 1* & 0 (0) & 0.071 & 1* (0) & 1* \\
          &       & 2B    & = K   & {0} & $ \neq\bm 0$ & $ \neq\bm 0$ & 1* (0) & 1* & 0.019 (0) & 0.047 & 1* (0) & 1* \\
          &       & {3A} & {< K} & {0} & $ \neq\bm 0$ & $ =\bm 0$ & 0.726 (0.122) & 0.900 & 0.009 (0) & 0.073 & 0.82 (0.129) & 0.925 \\
          &       & {3B} & = K   & {0} & $ \neq\bm 0$ & $ =\bm 0$ & 0.563 (0.183) & 0.847 & 0.013 (0) & 0.055 & 0.729 (0.205) & 0.916 \\
          &       & {4A} & {< K} & {0} & $ =\bm 0$ & $ \neq\bm 0$ & 0.018* (0.019) & 0.049* & 0.019 (0) & 0.062 & 0.016* (0.089) & 0.050* \\
          &       & {4B} & = K   & {0} & $ =\bm 0$ & $ \neq\bm 0$ & 0.004* (0.011) & 0.056* & 0.009 (0) & 0.049 & 0.006* (0.058) & 0.056* \\
          &       & 5     & < K   & $ \neq 0$    & $ =\bm 0$ & $ \neq\bm 0$ & 1* (0) & 1* & 1* (0) & 1* & 0.006* (0.055) & 0.052* \\
          &       & 6     & < K   & $ \neq 0$    & $ \neq\bm 0$ & $ =\bm 0$ & 1* (0) & 1* & 1* (0) & 1* & 0.813 (0.146) & 0.955 \\
    \bottomrule
    \end{tabular}%
    }\caption{Size and power $(^*)$ for every DGP, model specification (conditional or unconditional) and type of test statistic.\\
    Values not in brackets represent the frequency of rejection of the null hypothesiS,
        values in brackets denote the relative amount of inconclusive tests with the bound PSS approach (or the procedure by \citet{sam2019augmented}, denoted by SMG).}
        \label{tab:sim1}
    \end{table}
    \newpage
 The flow-chart in Figure \ref{fig:flowchart} depicts the possible outcomes of the analysis, based on each test outcome. In general, the $F_{ov}$ test, either bootstrap or ARDL, fails to correctly dismiss the presence of cointegrating relationships very often when degenerate cases occur. The complementary $t$ and $F_{ind}$ tests come to aid in these situations. The bootstrap $t$ test exhibits a good performance in terms of coverage and power, while the bound version, available only for case III, signals the presence of a degenerate case of first type with a rate that does not match the nominal coverage. 
    In particular, the bootstrap $F_{ov}$ test fails very often in both the conditional and unconditional model (it rejects the null too often, except for the DGP 3 in the conditional model), while the $t$ test leads to the right conclusion almost always with a coverage that is slightly better when the explanatory variables $\bm{x}_{t}$ are cointegrated rather than stationary. Thus, in case of rejection of the conditional bootstrap $F_{ov}$ test, if the $t$ test accepts the null, then the hypothesis of cointegration can be dismissed (DGPs 2 and 4). Otherwise, if the bootstrap $t$ test rejects the null, then a comparison of the results provided by the bootstrap $F_{ind}$ test in the conditional and unconditional model can help to disentangle between the other degenerate cases. DGP 5 or 6 are in place if the results of these tests are discordant.\\
  Finally, Figures \ref{fig:boot1} and \ref{fig:boot2} show examples of the kernel densities of the bootstrap distributions of  $F_{ov}^{H_0}$, $t^{H_0}$ and $F_{ind}^{H_0}$ statistics, for cases II and III respectively, under different DGPs.
  It is worth noting the difference in shape between the conditional and unconditional distributions under the null of the tests, especially for case II, as a possible result of the misspecification issue affecting the UC model. This is particularly true for DGP 1, that  better visualizes the differences in shape of the null distributions between the C and UC models than other DGPs, especially in the setting where explanatory variables are cointegrated.
    \newpage
\tikzset{
  treenode/.style = {shape=rectangle, rounded corners,
                     draw,align=center,
                     top color=white,
                     inner sep=4ex},
  decision/.style = {treenode, diamond, inner sep=3pt},
  root/.style     = {treenode},
  env/.style      = {treenode},
  ginish/.style   = {root},
  dummy/.style    = {circle,draw}
}
\newcommand{\yes}{edge node [above=0.2cm] {yes}}
\newcommand{\yesx}{edge  node [right=0.2cm]  {yes}}
\newcommand{\yesy}{edge  node [left=0.2cm]  {yes}}
\newcommand{\yesyc}{edge  node [above=0.2cm]  {yes}}
\newcommand{\no}{edge  node [left=0.2cm]  {no}}
\newcommand{\nox}{edge  node [above=0.2cm]  {no}}
\newcommand{\noy}{edge  node [right=0.2cm]  {no}}
\newcommand{\noyc}{edge  node [below=0.2cm]  {no}}
\newcommand{\noyb}{edge  node [above=0.2cm]  {no}}
\newcommand{\noz}{edge  node [below=0.2cm]  {no}}

\begin{landscape}
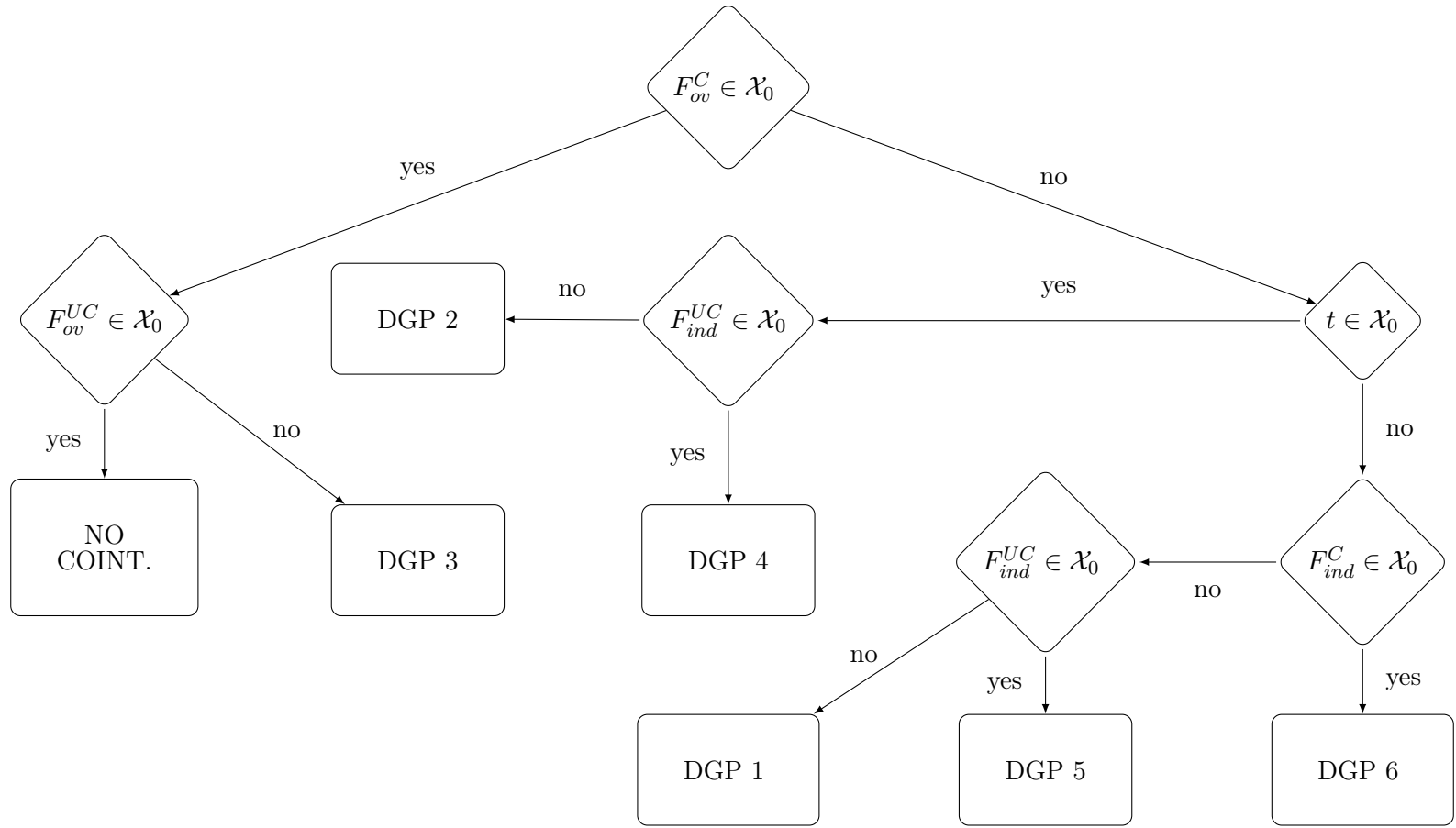
\begin{figure}[ht!]
\centering
\begin{tikzpicture}[-latex][scale=0.50]
  \matrix (chart)
    [
      matrix of nodes,
      column sep      = 5em,
      row sep         = 5ex,
      column 1/.style = {nodes={decision}},
      row 2/.style = {nodes={decision}},
      row 5/.style = {nodes={env}}
    ]
    {
       &
       &
       |[decision]|$F^{C}_{ov}\in \mathcal X_0$ &
       &
       \\
       |[decision]|$F^{UC}_{ov}\in \mathcal X_0$&
       |[treenode]|DGP 2&
       |[decision]|$F^{UC}_{ind}\in \mathcal X_0$&
       &
       |[decision]|$t\in \mathcal X_0$\\
       |[treenode]|\shortstack{NO \\ COINT.}&
       |[treenode]|DGP 3&
       |[treenode]|DGP 4&
       |[decision]| $F^{UC}_{ind}\in \mathcal X_0$ &
       |[decision]|$F^{C}_{ind}\in \mathcal X_0$\\
       &
       &|[treenode]| DGP 1
       &
       |[treenode]| DGP 5&
       |[treenode]| DGP 6
       \\};
       \draw
		 (chart-1-3) \yesyc (chart-2-1);
		 \draw
		 (chart-1-3) \noyb (chart-2-5);
		 \draw
		 (chart-2-5) \yes (chart-2-3);
		 \draw
		 (chart-2-3) \nox (chart-2-2);
		\draw
		 (chart-2-3)\yesy (chart-3-3);
		 \draw
		 (chart-2-5) \noy (chart-3-5);
		 \draw
		 (chart-3-5) \noyc (chart-3-4);
		 \draw
		 (chart-3-5) \yesx (chart-4-5);
		 \draw
		 (chart-3-4) \no (chart-4-3);
		 \draw
		 (chart-3-4) \yesy (chart-4-4);
		 \draw
		 (chart-2-1) \yesy (chart-3-1);
		 \draw
		 (chart-2-1) \noy (chart-3-2);
		 
  \end{tikzpicture}
	\caption{Flow-chart of the ARDL bootstrap cointegration tests. $\mathcal X_0$ denotes the acceptance region of the generic test statistic,$F^{C}_{.}$ and $F^{UC}_{.}$ denotes the $F_{.}$ test in the conditional and unconditional model, respectively.}\label{fig:flowchart}
	\end{figure}
\end{landscape}
\newpage

    \section{The ARDL Bootstrap Testing at Work}\label{sec:4sec}
\begin{figure}
        \centering
        \includegraphics[scale=0.65]{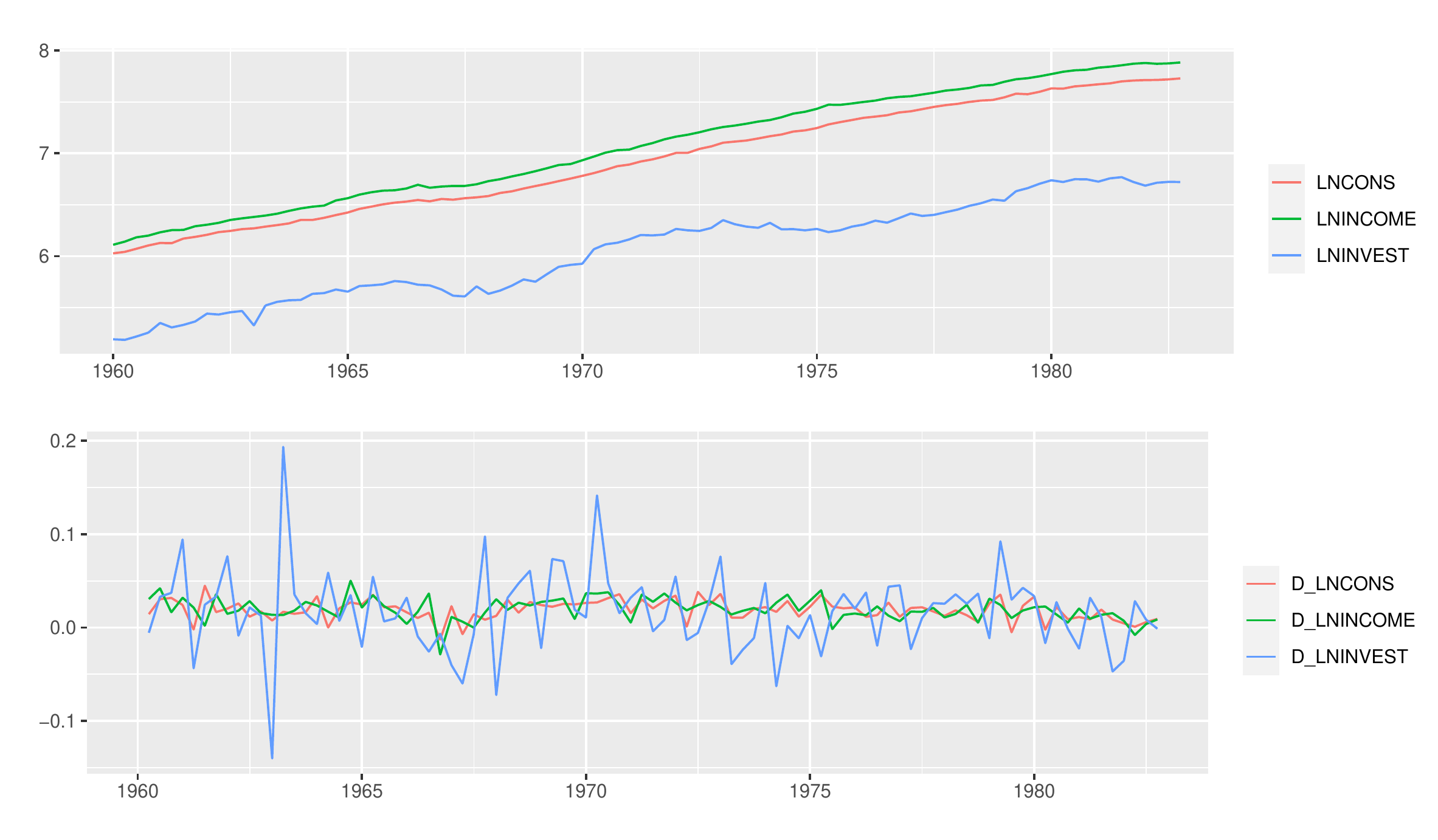}
        \caption{log-consumption/investment/income graphs (level variables and first differences).}
        \label{fig:plotemp}
    \end{figure}

    This section provides two illustrative applications which highlight the performance of the bootstrap ARDL tests. Both the applications have been carried out by using the R software.\\ In the first application we have examined the long-run relationship between consumption [C], income [INC], and investment [INV] of Germany via an ARDL model where consumption is the dependent variable.
   The model has been estimated by employing the dataset of \citet{lutkepohl2005} which includes quarterly data of the series over the years 1960 to 1982.
    The data have been employed in logarithmic form. Figure \ref{fig:plotemp} displays these series over the sample period.\\
    Before applying the bootstrap procedure, the order of integration of each series has been analyzed. Table \ref{tab:adf} shows the results of ADF test performed  on  both the series and their first-difference ($k=3$ maximum lags). They  confirm the applicability of the ARDL framework as no series is integrated of order higher than one.\\

    \begin{center}
\begin{table}[htbp]
\centering
  \resizebox{0.5\textwidth}{!}{
    \begin{tabular}{crrrrr}
          &       & \multicolumn{2}{c}{level variable} & \multicolumn{2}{c}{first difference} \\
\cmidrule{3-6}    Series & \multicolumn{1}{c}{lag} & \multicolumn{1}{c}{ADF} & \multicolumn{1}{c}{p.value} & \multicolumn{1}{c}{ADF} & \multicolumn{1}{c}{p-value} \\
    \midrule
    \multirow{4}[2]{*}{$\log\text{C}_t$} 
          & 0     & -1.690 & 0.450 & -9.750 & $< 0.01$ \\
          & 1     & -1.860 & 0.385 & -5.190 & $< 0.01$ \\
          & 2     & -1.420 & 0.549 & -3.130 & 0.030 \\
          & 3     & -1.010 & 0.691 & -2.720 & 0.080 \\
    \midrule
    \multirow{4}[2]{*}{$\log\text{INV}_t$} 
          & 0     & -2.290 & 0.217 & -11.140 & $<0.01$ \\
          & 1     & -1.960 & 0.345 & -7.510 & $< 0.01$ \\
          & 2     & -1.490 & 0.524 & -5.120 & $< 0.01$ \\
          & 3     & -1.310 & 0.587 & -3.290 & 0.020 \\
    \midrule
    \multirow{4}[2]{*}{$\log\text{INC}_t$} 
          & 0     & -1.200 & 0.625 & -8.390 & $< 0.01$ \\
          & 1     & -1.370 & 0.565 & -5.570 & $< 0.01$ \\
          & 2     & -1.360 & 0.570 & -3.300 & 0.020 \\
          & 3     & -1.220 & 0.619 & -3.100 & 0.032 \\
    \bottomrule
    \end{tabular}
  }
  \caption{ADF preliminary test (null hypothesis: random walk with drift).}
  \label{tab:adf}
\end{table}%
\end{center}
     The ARDL equation explaining the log consumption as function of the other variables is
        \begin{align}
    \Delta \log \text{C}_t &= \alpha_{0.y} -
    a_{yy} \log \text{C}_{t-1} - {a}_{y.x_1}\log \text{INV}_{t-1} - {a}_{y.x_2}\log \text{INC}_{t-1} +\\\nonumber
    &\sum_{j=1}^{p-1}\gamma_{y.j} \Delta\log \text{C}_{t-j} +
    \sum_{j=1}^{s-1}\gamma_{x_1.j} \Delta\log \text{INV}_{t-j} +
    \sum_{j=1}^{r-1}\gamma_{x_2.j} \Delta\log \text{INC}_{t-j} +\\\nonumber
    &\omega_1 \Delta\log \text{INV}_{t}+
    \omega_2 \Delta\log \text{INC}_{t}+\nu_{t} 
    \end{align}
    while the VECM marginal model explaining the other variables is composed of the following two equations
  \begin{align}
    \Delta \log \text{INV}_t &= \alpha_{0.x_{1}} 
     - {a}_{x_1,x_1}\log \text{INV}_{t-1} - {a}_{x_1,x_2}\log \text{INC}_{t-1} +\nonumber\\
    &\sum_{j=1}^{p-1}\gamma_{x_1,y,j} \Delta\log \text{C}_{t-j} +
    \sum_{j=1}^{s-1}\gamma_{x_1,x_1,j} \Delta\log \text{INV}_{t-j} +
    \sum_{j=1}^{r-1}\gamma_{x_1,x_2,j} \Delta\log \text{INC}_{t-j} +\varepsilon_{1,t} \nonumber\\
    \Delta \log \text{INC}_t &= \alpha_{0.x_{2}} 
     - {a}_{x_2,x_2}\log \text{INC}_{t-1}- {a}_{x_2,x_1}\log \text{INV}_{t-1}  +\nonumber\\
    &\sum_{j=1}^{p-1}\gamma_{x_2,y,j} \Delta\log \text{C}_{t-j} +
    \sum_{j=1}^{s-1}\gamma_{x_2,x_1,j} \Delta\log \text{INV}_{t-j} +
    \sum_{j=1}^{r-1}\gamma_{x_2,x_2,j} \Delta\log \text{INC}_{t-j} +\varepsilon_{2,t} \nonumber
    \end{align}   
The package \texttt{ARDL} has been employed to select the best number of lags (of the first differences) in the conditional ARDL equation, while the package \texttt{vars} has been used to determine the best number of lags in the VECM marginal model \\
    Table \ref{tab:resh1} reports the results of the best ARDL and VECM models. The results highlights that the dependent variable log-consumption is non stationary, which paves the way for the existence of a possible cointegrating relationship with the other variables.
    As for the latter, the Johansen test confirms the presence of a cointegrating relationship between them (see Table \ref{tab:resh1}).\\
    With the purpose of investigating the existence of a long-run relationship between log-consumption and the other variables, bound and bootstrap tests have been implemented, for both cases II and III. Table \ref{tab:test} shows the test results, confirming the existence of this long run relationship. \\
    The analysis has been repeated over a subset of the entire sample, from January, 1971 to December, 1982 ($T = 56$ observations), to test the effectiveness of the procedure with a reduced sample size, highlighting possible shortcomings of the PSS bound tests.\\ Table \ref{tab:resh2} shows the estimates for the best ARDL and VECM models. The Johansen test confirms the presence of a cointegrating relationship between $\log\text{INV}_t$ and $\log\text{INC}_t$ also in this subset of data, albeit with a slightly smaller test statistic.
    It is worth noting that the term $\Delta \bm x_t$, which appears in the model due to conditioning of $y_{t}$ on $\bm{x}_{t}$ is highly significant. Thus, omitting it may lead to bias in the estimates, and therefore to incorrect inference.\\
    Indeed, according to Table \ref{tab:test2}, which provides the test results, the existence of a long-run cointegrating relationship between the dependent and the independent variables is confirmed in the conditional ARDL, not in the unconditional one. Indeed, in this model, even if the $F_{ov}$ test rejects the null hypothesis, the $t$ and $F_{ind}$ tests state otherwise.\\ Eventually, it is worth noting that the asymptotic PSS and SMG tests are inconclusive on the matter. This allows us to conclude that the inclusion of instantaneous differences in the ARDL equation is a prominent requirement to allow for accurate inference of a possible cointegrating relationship among economic variables, and that the bootstrap tests herein proposed are a valid alternative in cases of uncertain asymptotic inference. However, the unconditional ARDL model, albeit suffering from misspecification, is still useful to detect degenerate cases of the second type, and should not be downright dismissed.
    
    \newpage
\begin{table}[ht!]
  \centering
  \resizebox{0.65\textwidth}{!}{
    \begin{tabular}{cllll}
          & \multicolumn{1}{c}{ARDL (C)} & \multicolumn{1}{c}{ARDL (UC)} & \multicolumn{2}{c}{VECM} \\
\cmidrule{2-5}          & \multicolumn{1}{c}{$\Delta\log\text{C}_t$} & \multicolumn{1}{c}{$\Delta\log\text{C}_t$} & \multicolumn{1}{c}{$\Delta\log\text{INV}_t$} & \multicolumn{1}{c}{$\Delta\log\text{INC}_t$} \\
    \midrule
    $\log\text{C}_{t-1}$ & \makecell{-0.307 ***\\ (0.055)} & \makecell{-0.316 ***\\ (0.07)} &       &  \\
    $\log\text{INV}_{t-1}$ & \makecell{-0.001\\ (0.011)} & \makecell{-0.001\\ (0.014)} & \makecell{-0.152 *\\ (0.063)} & \makecell{0.016\\ (0.017)} \\
    $\log\text{INC}_{t-1}$ & \makecell{0.297 ***\\ (0.055)} & \makecell{0.302 ***\\ (0.07)} & \makecell{0.124 *\\ (0.054)} & \makecell{-0.017\\ (0.014)} \\
    \midrule
    $\Delta\log\text{C}_{t-1}$ & \makecell{-0.248 **\\ (0.079)} & \makecell{-0.103\\ (0.098)} & \makecell{0.899 *\\ (0.442)} & \makecell{0.211 .\\ (0.113)} \\
    $\Delta\log\text{C}_{t-2}$ &       &       & \makecell{0.744 \\ (0.431)} &  \\
    $\Delta\log\text{INV}_{t-1}$ &       &       & \makecell{-0.18\\ (0.111)} & \makecell{0.035\\ (0.029)} \\
    $\Delta\log\text{INV}_{t-2}$ &       & \makecell{0.05\\ (0.027) }&       & \makecell{0.049 .\\ (0.027) }\\
    \midrule
    $\Delta\log\text{INV}_t$ & \makecell{0.065 **\\ (0.019)} &       &       &  \\
    $\Delta\log\text{INC}_t$ & \makecell{0.471 ***\\ (0.074)} &       &       &  \\
    const. & \makecell{0.048 ***\\ (0.013) }& \makecell{0.074 ***\\ (0.016)} & \makecell{0.036\\ (0.066)} & \makecell{0.033 *\\ (0.017)} \\
    \hline
     $rk(\bm A_{xx})\leq 1$&\multicolumn{2}{c}{$J_T=6.16$}&\multicolumn{2}{c}{$J_E=6.16$}\\
     $rk(\bm A_{xx})=0$&\multicolumn{2}{c}{$J_T=51.04$***}&\multicolumn{2}{c}{$J_E=44.89$***}\\
    \bottomrule
    \end{tabular}%
    }

    \caption{ARDL and VECM results for the consumption/investment/income dataset (whole sample).\\
    The last two rows report the Johansen test on the marginal VECM model, based on the trace ($J_T$) and largest eigenvalue ($J_E$) statistics.\\
    Significance codes: (***) 1\%; (**) 5\%; (.) 10\%.}
  \label{tab:resh1}%
\end{table}%

\begin{table}[h!]
  \centering\resizebox{0.8\textwidth}{!}{
    \begin{tabular}{clrrrrrr}
          &       & \multicolumn{1}{c}{$F_{ov}$ (C)} & \multicolumn{1}{c}{$F_{ov}$ (UC)} & \multicolumn{1}{c}{$t$ (C)} & \multicolumn{1}{c}{$t$ (UC)} & \multicolumn{1}{c}{$F_{ind}$ (C)} & \multicolumn{1}{c}{$F_{ind}$ (UC)} \\
    \midrule
    \multirow{4}[2]{*}{Case II} & Statistic & 18.019 & 27.835 & -5.608 & -4.49 & 15.636 & 9.879 \\
          & I(0) 5\% & 3.1   & 3.1   & \multicolumn{1}{l}{-} & \multicolumn{1}{l}{-} & \multicolumn{1}{l}{-} & \multicolumn{1}{l}{-} \\
          & I(1) 5\%& 3.87  & 3.87  & \multicolumn{1}{l}{-} & \multicolumn{1}{l}{-} & \multicolumn{1}{l}{-} & \multicolumn{1}{l}{-} \\
          & Boot. p-value & 0.0005 & 0.0005 & 0.0005 & 0.0005 & 0.0005 & 0.0005 \\
    \midrule
    \multirow{4}[2]{*}{Case III} & Statistic & 10.751 & 7.967 & -5.608 & -4.49 & 15.636 & 9.879 \\
          & I(0) 5\% & 3.79  & 3.79  & -2.86 & -2.86 &   3.01    & 3.01 \\
          & I(1) 5\% & 4.85  & 4.85  & -3.53 & -3.53 &   5.42    & 5.42 \\
          & Boot. p-value & 0.0005 & 0.001 & 0.0005 & 0.0005 & 0.0005 & 0.001 \\
    \bottomrule
    \end{tabular}%
    }
  \caption{ARDL cointegration test results for the consumption/investment/income dataset (whole sample).}
  \label{tab:test}%
\end{table}%
\newpage
\begin{table}[ht!]
  \centering
   \resizebox{0.6\textwidth}{!}{
    \begin{tabular}{cllll}
  
          & \multicolumn{1}{c}{ARDL (C)} & \multicolumn{1}{c}{ARDL (UC)} & \multicolumn{2}{c}{VECM} \\
\cmidrule{2-5}          & \multicolumn{1}{c}{$\Delta\log\text{C}_t$} & \multicolumn{1}{c}{$\Delta\log\text{C}_t$} & \multicolumn{1}{c}{$\Delta\log\text{INV}_t$} & \multicolumn{1}{c}{$\Delta\log\text{INC}_t$} \\
    \midrule
    $\log\text{C}_{t-1}$ & \makecell{-0.254 **\\ (0.087) }& \makecell{-0.177\\(0.126) }&       &  \\
    $\log\text{INV}_{t-1}$ & \makecell{-0.007  \\(0.013) }& \makecell{-0.014\\ (0.019)} & \makecell{-0.099 \\ (0.061)} & \makecell{0.014\\ (0.016)} \\
    $\log\text{INC}_{t-1}$ & \makecell{0.274 **\\ (0.089) }& \makecell{0.167 \\(0.127) }& \makecell{0.10 . \\(0.054) }& \makecell{-0.03 *\\ (0.014)} \\
    \midrule
    $\Delta\log\text{C}_{t-1}$ & \makecell{-0.218 * \\(0.112) }& \makecell{-0.257 . \\(0.149)} &       &  \\
    \midrule
    $\Delta\log\text{INV}_t$ & \makecell{0.146 *** \\(0.032)} &       &       &  \\
    $\Delta\log\text{INC}_t$ & \makecell{0.674 *** \\(0.123)} &       &       &  \\
    const. & \makecell{0.015 \\(0.055) }& \makecell{0.164 *\\ (0.07) }& \makecell{-0.118 \\(0.19)} & \makecell{0.157 ** \\(0.043)} \\
    \hline
     $rk(\bm A_{xx})\leq 1$&\multicolumn{2}{c}{$J_T=3.73$}&\multicolumn{2}{c}{$J_E=3.73$}\\
     $rk(\bm A_{xx})=0$&\multicolumn{2}{c}{$J_T=30.26$***}&\multicolumn{2}{c}{$J_E=26.53$***}\\
    \bottomrule
    \end{tabular}%
    }
    \caption{ARDL and VECM results for the consumption/investment/income dataset (from 1971).\\
     The last two rows report the Johansen test on the marginal VECM model, based on the trace ($J_T$) and largest eigenvalue ($J_E$) statistics.\\
    Significance codes: (***) 1\%; (**) 5\%; (.) 10\%.}
  \label{tab:resh2}%
\end{table}%

\begin{table}[htbp!]
  \centering\resizebox{0.8\textwidth}{!}{
    \begin{tabular}{clrrrrrr}
          &       & \multicolumn{1}{c}{$F_{ov}$ (C)} & \multicolumn{1}{c}{$F_{ov}$ (UC)} & \multicolumn{1}{c}{$t$ (C)} & \multicolumn{1}{c}{$t$ (UC)} & \multicolumn{1}{c}{$F_{ind}$ (C)} & \multicolumn{1}{c}{$F_{ind}$ (UC)} \\
    \midrule
    \multirow{4}[2]{*}{Case II} & Statistic & 5.942 & 5.683 & -3.112 & -1.404 & 5.014 & 1.288 \\
          & I(0) 5\% & 3.435 & 3.435 & \multicolumn{1}{l}{-} & \multicolumn{1}{l}{-} & \multicolumn{1}{l}{-} & \multicolumn{1}{l}{-} \\
          & I(1) 5\%& 4.26  & 4.26  & \multicolumn{1}{l}{-} & \multicolumn{1}{l}{-} & \multicolumn{1}{l}{-} & \multicolumn{1}{l}{-} \\
          & Boot. p-value & 0.044 & 0.0001 & 0.005 & 0.345 & 0.032 & 0.782 \\
    \midrule
    \multirow{4}[2]{*}{Case III} & Statistic & 5.942 & 5.683 & -3.112 & -1.404 & 5.014 & 1.288 \\
          & I(0) 5\% & 4.133 & 4.133 & -2.86 & -2.86 &   3.22    & 3.22 \\
          & I(1) 5\% & 5.26  & 5.26  & -3.53 & -3.53 &   5.62    & 5.62 \\
          & Boot. p-value & 0.026 & 0.008  & 0.004 & 0.341 & 0.032 & 0.786 \\
    \bottomrule
    \end{tabular}%
    }
  \caption{ARDL cointegration test results for the consumption/investment/income dataset (from 1971).}
  \label{tab:test2}%
\end{table}%
\newpage
As a second application, following  \citet{McNown2017}, we have investigated the relationship between foreign direct investment [FDI], exports [EXP], and gross domestic product [GDP] in some OECD economies using the bootstrap ARDL test for cointegration. The Countries considered for this analysis are: Germany (DE), France (FR), Spain (ES), England (UK) and Italy (IT).
The data of these three yearly variables have been retrieved from the World Bank Database and cover the period from 1960 to 2020. In the analysis, the logarithm form of the variables has been used and [EXP] and [FDI] have been adjusted using the GDP deflator. 
Table \ref{tab:gdp1} shows the outcomes of the ADF test performed on each variable, which ensure that the integration order is not higher than one for all variables. Table \ref{tab:cointbig} shows the results of bound and bootstrap tests performed in both the conditional (C) and the unconditional (UC) ARDL model by taking each variable, in turn, as the dependent one.
The following ARDL equations have been estimated for each of the five aforementioned countries. For the sake of simplicity, we have omitted the VECM marginal models pertaining to the explanatory variables of each cointegrating analysis and each country.
\begin{itemize}
    \item First ARDL equation:
    \begin{align}
    \Delta \log \text{GDP}_{t}&=\alpha_{0.y} -
    a_{yy} \log \text{GDP}_{t-1} - {a}_{y.x_1}\log \text{EXP}_{t-1} - {a}_{y.x_2}\log \text{FDI}_{t-1} +\\\nonumber
    &\sum_{j=1}^{p-1}\gamma_{y.j} \Delta\log \text{GDP}_{t-j} +
    \sum_{j=1}^{s-1}\gamma_{x_1.j} \Delta\log \text{EXP}_{t-j} +
    \sum_{j=1}^{r-1}\gamma_{x_2.j} \Delta\log \text{FDI}_{t-j} +\\\nonumber
    &\omega_1 \Delta\log \text{EXP}_{t}+
    \omega_2 \Delta\log \text{FDI}_{t}+\nu_{t} 
    \end{align}
    For this model, the long-run cointegrating relationship is present only for Germany, while a degenerate case of the second type can be observed for Italy. Again, an analysis based on the UC ARDL model (i.e., disregarding $\Delta\log \text{EXP}$ and $\Delta\log \text{FDI}$) leads to results that differ from those obtained from the conditional ARDL model for Germany and Italy. Furthermore, the bound testing procedure does not always lead to conclusive results. This happens for the conditional model for Germany and for the unconditional model for Italy. 
    \item Second ARDL equation:
    \begin{align}
    \Delta \log \text{EXP}_{t}&=\alpha_{0.y} -
    a_{yy} \log \text{EXP}_{t-1} - {a}_{y.x_1}\log \text{GDP}_{t-1} - {a}_{y.x_2}\log \text{FDI}_{t-1} +\\\nonumber
    &\sum_{j=1}^{p-1}\gamma_{y.j} \Delta\log \text{EXP}_{t-j} +
    \sum_{j=1}^{s-1}\gamma_{x_1.j} \Delta\log \text{GDP}_{t-j} +
    \sum_{j=1}^{r-1}\gamma_{x_2.j} \Delta\log \text{FDI}_{t-j} +\\\nonumber
    &\omega_1 \Delta\log \text{GDP}_{t}+
    \omega_2 \Delta\log \text{FDI}_{t}+\nu_{t} 
    \end{align}
    For this model, the long-run cointegrating relationship is present for Spain and France, while a degenerate case of the second type can be observed for Italy. An analysis based on the UC model (i.e., disregarding $\Delta\log \text{GDP}$ and $\Delta\log \text{FDI}$) produces results different from those obtained from the conditional ARDL model for France. Furthermore, the bound testing approach is inconclusive for France (conditional model) and Italy (unconditional model).
    \item Third ARDL equation:
    \begin{align}
    \Delta \log \text{FDI}_{t}&=\alpha_{0.y} -
    a_{yy} \log \text{FDI}_{t-1} - {a}_{y.x_1}\log \text{GDP}_{t-1} - {a}_{y.x_2}\log \text{EXP}_{t-1} +\\\nonumber
    &\sum_{j=1}^{p-1}\gamma_{y.j} \Delta\log \text{FDI}_{t-j} +
    \sum_{j=1}^{s-1}\gamma_{x_1.j} \Delta\log \text{GDP}_{t-j} +
    \sum_{j=1}^{r-1}\gamma_{x_2.j} \Delta\log \text{EXP}_{t-j} +\\\nonumber
    &\omega_1 \Delta\log \text{GDP}_{t}+
    \omega_2 \Delta\log \text{EXP}_{t}+\nu_{t} 
    \end{align}
        For this model, the long-run cointegrating relationship is present for all Countries except UK, while a degenerate case of the second type can be observed for Spain. The UC model (i.e., disregarding $\Delta\log \text{GDP}$ and $\Delta\log \text{EXP}$ leads to drastically different results than the conditional ARDL, confirming a cointegrating relationship only for Italy. The bound testing approach is inconclusive for the conditional models of all Countries except UK and Italy, while it generally mirrors the bootstrap version in the unconditional model, having no uncertain outcomes.
\end{itemize}

This empirical application further highlights the importance of dealing with inconclusive inference via the bootstrap procedure, while also incorporating the effect of conditioning in the ARDL model, since otherwise the inference may lead to an inaccurate conclusion, due to the omission of relevant predictors from the equation, in particular in the error correction term of the ARDL equation.
\newpage
\begin{table}[htbp]
  \centering
  \resizebox{\textwidth}{!}{
    \begin{tabular}{clrrrrrrrrrrrr}
          &       & \multicolumn{4}{c}{No Drift, No Trend} & \multicolumn{4}{c}{Drift, No Trend} & \multicolumn{4}{c}{Drift and Trend} \\
\cmidrule{3-14}    \multicolumn{1}{l}{Countries} & Variable & \multicolumn{1}{l}{Lag = 0} & \multicolumn{1}{l}{Lag = 1} & \multicolumn{1}{l}{Lag = 2} & \multicolumn{1}{l}{Lag = 3} & \multicolumn{1}{l}{Lag = 0} & \multicolumn{1}{l}{Lag = 1} & \multicolumn{1}{l}{Lag = 2} & \multicolumn{1}{l}{Lag = 3} & \multicolumn{1}{l}{Lag = 0} & \multicolumn{1}{l}{Lag = 1} & \multicolumn{1}{l}{Lag = 2} & \multicolumn{1}{l}{Lag = 3} \\
    \midrule
    \multirow{6}[4]{*}{DE} &
    $\log \text{GDP}_t$  & 0.99  & 0.99  & 0.99  & 0.99  & 0.0962 & 0.2287 & 0.1661 & 0.2342 & 0.902 & 0.841 & 0.954 & 0.978 \\
    & $\log \text{FDI}_t$   & 0.844 & 0.906 & 0.857 & 0.921 & 0.4   & 0.544 & 0.507 & 0.56  & 0.0445 & 0.2489 & 0.0485 & 0.2093 \\
          & $\log \text{EXP}_t$  & 0.99  & 0.982 & 0.987 & 0.976 & 0.108 & 0.287 & 0.461 & 0.577 & 0.411 & 0.15  & 0.502 & 0.486 \\
\cmidrule{2-14}          & $\Delta\log \text{GDP}_t$ & $<0.01$  & $<0.01$  & 0.0163 & 0.0709 & $<0.01$  & $<0.01$  & $<0.01$  & 0.037 & $<0.01$  & $<0.01$  & $<0.01$  & 0.0181 \\
          & $\Delta\log \text{FDI}_t$  & $<0.01$  & $<0.01$  & $<0.01$  & $<0.01$  & $<0.01$  & $<0.01$  & $<0.01$  & $<0.01$  & $<0.01$  & $<0.01$  & $<0.01$  & $<0.01$ \\
          & $\Delta\log \text{EXP}_t$ & $<0.01$  & $<0.01$  & $<0.01$  & 0.014 & $<0.01$  & $<0.01$  & $<0.01$  & 0.0457 & $<0.01$  & $<0.01$  & 0.0218 & 0.1014 \\
    \midrule
    \multirow{6}[4]{*}{ES} & $\log \text{GDP}_t$  & 0.99  & 0.854 & 0.901 & 0.911 & 0.0472 & 0.3495 & 0.4678 & 0.5065 & 0.99  & 0.973 & 0.99  & 0.99 \\
          & $\log \text{FDI}_t$   & 0.197 & 0.443 & 0.55  & 0.597 & $<0.01$  & $<0.01$  & 0.0226 & 0.0569 & $<0.01$  & $<0.01$  & 0.0375 & 0.0894 \\
          & $\log \text{EXP}_t$  & 0.952 & 0.923 & 0.966 & 0.945 & 0.842 & 0.695 & 0.802 & 0.796 & 0.423 & 0.404 & 0.571 & 0.454 \\
\cmidrule{2-14}          & $\Delta\log \text{GDP}_t$ & 0.055 & 0.0262 & 0.031 & 0.045 & 0.378 & 0.263 & 0.271 & 0.301 & 0.399 & 0.306 & 0.323 & 0.338 \\
          & $\Delta\log \text{FDI}_t$  & $<0.01$  & $<0.01$  & $<0.01$  & $<0.01$  & $<0.01$  & $<0.01$  & $<0.01$  & $<0.01$  & $<0.01$  & $<0.01$  & $<0.01$  & $<0.01$ \\
          & $\Delta\log \text{EXP}_t$ & $<0.01$  & $<0.01$  & $<0.01$  & $<0.01$  & $<0.01$  & $<0.01$  & 0.0175 & 0.0433 & $<0.01$  & $<0.01$  & 0.0829 & 0.1961 \\
    \midrule
    \multirow{6}[4]{*}{FR} & $\log \text{GDP}_t$  & 0.99  & 0.987 & 0.979 & 0.978 & $<0.01$  & 0.0582 & 0.0516 & 0.0824 & 0.975 & 0.989 & 0.99  & 0.99 \\
          & $\log \text{FDI}_t$   & 0.837 & 0.882 & 0.817 & 0.827 & 0.378 & 0.363 & 0.393 & 0.507 & 0.499 & 0.603 & 0.527 & 0.702 \\
          & $\log \text{EXP}_t$  & 0.946 & 0.881 & 0.909 & 0.875 & 0.689 & 0.54  & 0.655 & 0.711 & 0.638 & 0.35  & 0.455 & 0.294 \\
\cmidrule{2-14}          & $\Delta\log \text{GDP}_t$ & 0.0102 & 0.0294 & 0.0262 & 0.0458 & 0.0234 & 0.078 & 0.0896 & 0.1846 & $<0.01$  & 0.0129 & 0.0149 & 0.0205 \\
          & $\Delta\log \text{FDI}_t$  & $<0.01$  & $<0.01$  & $<0.01$  & $<0.01$  & $<0.01$  & $<0.01$  & $<0.01$  & $<0.01$  & $<0.01$  & $<0.01$  & $<0.01$  & 0.0176 \\
          & $\Delta\log \text{EXP}_t$ & $<0.01$  & $<0.01$  & $<0.01$  & $<0.01$  & $<0.01$  & $<0.01$  & 0.0167 & 0.017 & $<0.01$  & $<0.01$  & 0.068 & 0.0686 \\
    \midrule
    \multirow{6}[4]{*}{UK} & $\log \text{GDP}_t$  & 0.99  & 0.982 & 0.99  & 0.983 & 0.392 & 0.488 & 0.524 & 0.63  & 0.99  & 0.972 & 0.99  & 0.99 \\
          &$\log \text{FDI}_t$  & 0.0325 & 0.0654 & 0.1604 & 0.2651 & $<0.01$  & $<0.01$  & 0.0525 & 0.3877 & $<0.01$  & $<0.01$  & 0.0137 & 0.1819 \\
          & $\log \text{EXP}_t$  & 0.9   & 0.781 & 0.858 & 0.833 & 0.58  & 0.509 & 0.586 & 0.699 & 0.725 & 0.464 & 0.664 & 0.652 \\
\cmidrule{2-14}          & $\Delta\log \text{GDP}_t$ & $<0.01$  & 0.012 & 0.0224 & 0.1452 & 0.0223 & 0.0219 & 0.0544 & 0.276 & 0.0489 & 0.0477 & 0.1481 & 0.3992 \\
          & $\Delta\log \text{FDI}_t$  & $<0.01$  & $<0.01$  & $<0.01$  & $<0.01$  & $<0.01$  & $<0.01$  & $<0.01$  & $<0.01$  & $<0.01$  & $<0.01$  & $<0.01$  & $<0.01$ \\
          & $\Delta\log \text{EXP}_t$ & $<0.01$  & $<0.01$  & $<0.01$  & $<0.01$  & $<0.01$  & $<0.01$  & $<0.01$  & $<0.01$  & $<0.01$  & $<0.01$  & $<0.01$  & 0.0192 \\
    \midrule
    \multirow{6}[4]{*}{IT} &$\log \text{GDP}_t$  & 0.99  & 0.974 & 0.941 & 0.796 & $<0.01$  & $<0.01$  & $<0.01$  & 0.084 & 0.99  & 0.99  & 0.99  & 0.99 \\
          & $\log \text{FDI}_t$   & 0.572 & 0.599 & 0.675 & 0.725 & $<0.01$  & 0.0759 & 0.3199 & 0.5174 & $<0.01$  & 0.013 & 0.151 & 0.46 \\
          & $\log \text{EXP}_t$  & 0.787 & 0.71  & 0.698 & 0.684 & 0.479 & 0.288 & 0.467 & 0.433 & 0.629 & 0.35  & 0.463 & 0.379 \\
\cmidrule{2-14}          & $\Delta\log \text{GDP}_t$ & $<0.01$  & $<0.01$64 & 0.0429 & 0.0402 & $<0.01$  & 0.0861 & 0.3989 & 0.4267 & $<0.01$  & $<0.01$  & 0.0166 & 0.017 \\
          & $\Delta\log \text{FDI}_t$  & $<0.01$  & $<0.01$  & $<0.01$  & $<0.01$  & $<0.01$  & $<0.01$  & $<0.01$  & $<0.01$  & $<0.01$  & $<0.01$  & $<0.01$  & $<0.01$ \\
          & $\Delta\log \text{EXP}_t$ & $<0.01$  & $<0.01$  & $<0.01$  & $<0.01$  & $<0.01$  & $<0.01$  & $<0.01$  & $<0.01$  & $<0.01$  & $<0.01$  & 0.0336 & 0.0315 \\
    \bottomrule
    \end{tabular}%
    }
  \caption{ADF preliminary test for the second example.}
  \label{tab:gdp1}%
\end{table}%
\newpage

\begin{table}[htbp]
  \centering
  \resizebox{\textwidth}{!}{
    \begin{tabular}{cccccccccccccc}
          &       &       &       & \multicolumn{2}{c}{Critical Values} & \multicolumn{2}{c}{PSS / SMG Threshold} & \multicolumn{2}{c}{Statistic} & \multicolumn{4}{c}{Outcome} \\
    \midrule
    \multicolumn{1}{c}{Model} & \multicolumn{1}{c}{Country} & \multicolumn{1}{c}{Lags} & Test  & \multicolumn{1}{c}{C} & \multicolumn{1}{c}{UC} & \multicolumn{1}{c}{I(0) 5\%} & \multicolumn{1}{c}{I(1) 5\%} & \multicolumn{1}{c}{C} & \multicolumn{1}{c}{UC} & \multicolumn{1}{c}{C (boot)} & \multicolumn{1}{c}{UC (boot)} & \multicolumn{1}{c}{C (bound)} & \multicolumn{1}{c}{UC (bound)} \\
    \midrule
    \multirow{15}[10]{*}{$\log \text{GDP}|\log \text{FDI},\log \text{EXP}$} & \multirow{3}[2]{*}{DE} & \multirow{3}[2]{*}{(1,1,1)} & $F_{ov}$  & 4.192 & 3.192 & 4.070 & 5.190 & 5.132 & 5.215 & \multirow{3}[2]{*}{Y} & \multirow{3}[2]{*}{Y} & \multirow{3}[2]{*}{U} & \multirow{3}[2]{*}{Y} \\
          &       &       & t     & -1.486 & -1.367 & -2.860 & -3.530 & -3.848 & -3.852 &       &       &       &  \\
          &       &       & $F_{ind}$ & 4.936 & 4.177 & 3.220 & 5.620 & 7.088 & 6.364 &       &       &       &  \\
\cmidrule{2-14}          & \multirow{3}[2]{*}{ES} & \multirow{3}[2]{*}{(1,0,0)} & $F_{ov}$  & 7.990 & 5.600 & 4.080 & 5.210 & 2.437 & 1.378 & \multirow{3}[2]{*}{N} & \multirow{3}[2]{*}{N} & \multirow{3}[2]{*}{N} & \multirow{3}[2]{*}{N} \\
          &       &       & t     & -1.671 & -2.082 & -2.860 & -3.530 & -1.306 & 0.406 &       &       &       &  \\
          &       &       & $F_{ind}$ & 4.316 & 3.838 & 3.210 & 5.620 & 1.108 & 0.523 &       &       &       &  \\
\cmidrule{2-14}          & \multirow{3}[2]{*}{FR} & \multirow{3}[2]{*}{(1,1,1)} & $F_{ov}$  & 4.134 & 4.701 & 4.080 & 5.210 & 3.421 & 2.149 & \multirow{3}[2]{*}{N} & \multirow{3}[2]{*}{N} & \multirow{3}[2]{*}{N} & \multirow{3}[2]{*}{N} \\
          &       &       & t     & -1.631 & -2.110 & -2.860 & -3.530 & -2.504 & -0.326 &       &       &       &  \\
          &       &       & $F_{ind}$ & 4.292 & 4.900 & 3.210 & 5.620 & 1.618 & 0.630 &       &       &       &  \\
\cmidrule{2-14}          & \multirow{3}[2]{*}{UK} & \multirow{3}[2]{*}{(1,0,1)} & $F_{ov}$  & 3.278 & 4.213 & 4.070 & 5.190 & 1.795 & 2.357 & \multirow{3}[2]{*}{N} & \multirow{3}[2]{*}{N} & \multirow{3}[2]{*}{N} & \multirow{3}[2]{*}{N} \\
          &       &       & t     & -1.717 & -1.288 & -2.860 & -3.530 & 0.063 & 0.767 &       &       &       &  \\
          &       &       & $F_{ind}$ & 3.704 & 3.374 & 3.220 & 5.620 & 1.312 & 2.430 &       &       &       &  \\
\cmidrule{2-14}          & \multirow{3}[2]{*}{IT} & \multirow{3}[2]{*}{(1,1,0)} & $F_{ov}$  & 3.632 & 3.385 & 4.070 & 5.190 & 9.758 & 7.976 & \multirow{3}[2]{*}{D2} & \multirow{3}[2]{*}{D1} & \multirow{3}[2]{*}{N} & \multirow{3}[2]{*}{U} \\
          &       &       & t     & -1.680 & -1.784 & -2.860 & -3.530 & -2.338 & -1.603 &       &       &       &  \\
          &       &       & $F_{ind}$ & 3.648 & 3.306 & 3.220 & 5.620 & 2.273 & 4.191 &       &       &       &  \\
    \midrule
    \multirow{15}[10]{*}{$\log \text{EXP}|\log \text{GDP},\log \text{FDI}$} & \multirow{3}[2]{*}{DE} & \multirow{3}[2]{*}{(2,0,0)} & $F_{ov}$  & 5.880 & 4.932 & 4.070 & 5.190 & 2.706 & 1.826 & \multirow{3}[2]{*}{N} & \multirow{3}[2]{*}{N} & \multirow{3}[2]{*}{N} & \multirow{3}[2]{*}{N} \\
          &       &       & t     & -3.647 & -2.337 & -2.860 & -3.530 & -2.563 & -0.832 &       &       &       &  \\
          &       &       & $F_{ind}$ & 7.286 & 3.856 & 3.220 & 5.620 & 2.546 & 0.083 &       &       &       &  \\
\cmidrule{2-14}          & \multirow{3}[2]{*}{ES} & \multirow{3}[2]{*}{(1,5,1)} & $F_{ov}$  & 5.000 & 5.190 & 4.080 & 5.210 & 6.510 & 6.191 & \multirow{3}[2]{*}{Y} & \multirow{3}[2]{*}{Y} & \multirow{3}[2]{*}{Y} & \multirow{3}[2]{*}{Y} \\
          &       &       & t     & -1.898 & -2.701 & -2.860 & -3.530 & -3.724 & -4.282 &       &       &       &  \\
          &       &       & $F_{ind}$ & 3.850 & 4.995 & 3.210 & 5.620 & 9.420 & 9.110 &       &       &       &  \\
\cmidrule{2-14}          & \multirow{3}[2]{*}{FR} & \multirow{3}[2]{*}{(2,0,2)} & $F_{ov}$  & 4.911 & 5.323 & 4.080 & 5.210 & 3.241 & 2.474 & \multirow{3}[2]{*}{Y} & \multirow{3}[2]{*}{D2} & \multirow{3}[2]{*}{U} & \multirow{3}[2]{*}{N} \\
          &       &       & t     & -1.659 & -1.316 & -2.860 & -3.530 & -2.588 & -2.633 &       &       &       &  \\
          &       &       & $F_{ind}$ & 3.696 & 3.254 & 3.210 & 5.620 & 4.717 & 2.437 &       &       &       &  \\
\cmidrule{2-14}          & \multirow{3}[2]{*}{UK} & \multirow{3}[2]{*}{(5,5,5)} & $F_{ov}$  & 4.354 & 4.560 & 4.070 & 5.190 & 1.612 & 1.372 & \multirow{3}[2]{*}{N} & \multirow{3}[2]{*}{N} & \multirow{3}[2]{*}{N} & \multirow{3}[2]{*}{N} \\
          &       &       & t     & -1.913 & -2.649 & -2.860 & -3.530 & -0.651 & -1.523 &       &       &       &  \\
          &       &       & $F_{ind}$ & 3.661 & 4.570 & 3.220 & 5.620 & 2.323 & 1.769 &       &       &       &  \\
\cmidrule{2-14}          & \multirow{3}[2]{*}{IT} & \multirow{3}[2]{*}{(1,0,0)} & $F_{ov}$  & 5.438 & 4.901 & 4.070 & 5.190 & 2.649 & 1.760 & \multirow{3}[2]{*}{D2} & \multirow{3}[2]{*}{D2} & \multirow{3}[2]{*}{U} & \multirow{3}[2]{*}{N} \\
          &       &       & t     & -1.830 & -1.620 & -2.860 & -3.530 & -1.849 & -2.211 &       &       &       &  \\
          &       &       & $F_{ind}$ & 3.513 & 3.206 & 3.220 & 5.620 & 3.481 & 2.425 &       &       &       &  \\
    \midrule
    \multirow{15}[10]{*}{$\log \text{FDI}|\log \text{GDP},\log \text{EXP}$} & \multirow{3}[2]{*}{DE} & \multirow{3}[2]{*}{(1,0,0)} & $F_{ov}$  & 5.500 & 5.561 & 4.070 & 5.190 & 3.245 & 2.310 & \multirow{3}[2]{*}{Y} & \multirow{3}[2]{*}{D2} & \multirow{3}[2]{*}{U} & \multirow{3}[2]{*}{N} \\
          &       &       & t     & -1.767 & -1.704 & -2.860 & -3.530 & -2.869 & -2.021 &       &       &       &  \\
          &       &       & $F_{ind}$ & 3.622 & 3.578 & 3.220 & 5.620 & 4.085 & 1.177 &       &       &       &  \\
\cmidrule{2-14}          & \multirow{3}[2]{*}{ES} & \multirow{3}[2]{*}{(4,2,5)} & $F_{ov}$  & 6.015 & 5.792 & 4.080 & 5.210 & 7.513 & 7.501 & \multirow{3}[2]{*}{D2} & \multirow{3}[2]{*}{D2} & \multirow{3}[2]{*}{U} & \multirow{3}[2]{*}{D2} \\
          &       &       & t     & -1.640 & -1.698 & -2.860 & -3.530 & -4.648 & -4.695 &       &       &       &  \\
          &       &       & $F_{ind}$ & 4.463 & 3.600 & 3.210 & 5.620 & 3.357 & 2.673 &       &       &       &  \\
\cmidrule{2-14}          & \multirow{3}[2]{*}{FR} & \multirow{3}[2]{*}{(1,0,1)} & $F_{ov}$  & 3.492 & 5.346 & 4.080 & 5.210 & 4.182 & 3.722 & \multirow{3}[2]{*}{Y} & \multirow{3}[2]{*}{N} & \multirow{3}[2]{*}{U} & \multirow{3}[2]{*}{N} \\
          &       &       & t     & -1.808 & -1.726 & -2.860 & -3.530 & -3.418 & -2.805 &       &       &       &  \\
          &       &       & $F_{ind}$ & 3.984 & 3.415 & 3.210 & 5.620 & 4.107 & 2.257 &       &       &       &  \\
\cmidrule{2-14}          & \multirow{3}[2]{*}{UK} & \multirow{3}[2]{*}{(1,1,0)} & $F_{ov}$  & 4.761 & 3.965 & 4.080 & 5.210 & 2.038 & 1.904 & \multirow{3}[2]{*}{N} & \multirow{3}[2]{*}{N} & \multirow{3}[2]{*}{N} & \multirow{3}[2]{*}{N} \\
          &       &       & t     & -2.260 & -2.581 & -2.860 & -3.530 & -0.943 & -0.666 &       &       &       &  \\
          &       &       & $F_{ind}$ & 4.695 & 4.986 & 3.210 & 5.620 & 2.885 & 2.250 &       &       &       &  \\
\cmidrule{2-14}          & \multirow{3}[2]{*}{IT} & \multirow{3}[2]{*}{(1,0,1)} & $F_{ov}$  & 5.560 & 5.368 & 4.070 & 5.190 & 6.716 & 6.997 & \multirow{3}[2]{*}{Y} & \multirow{3}[2]{*}{Y} & \multirow{3}[2]{*}{Y} & \multirow{3}[2]{*}{Y} \\
          &       &       & t     & -1.656 & -1.865 & -2.860 & -3.530 & -4.202 & -4.507 &       &       &       &  \\
          &       &       & $F_{ind}$ & 4.464 & 3.241 & 3.220 & 5.620 & 7.017 & 5.882 &       &       &       &  \\
    \bottomrule
    \end{tabular}%
    }
  \caption{Cointegration analysis for the second example, divided by model equation and Country. The optimal number of ARDL lags in the short-run, bootstrap critical values, bound test thresholds and test statistics for each test are shown (case III).\\ The outcome columns draw conclusions on each procedure and each type of model (conditional or unconditional, bootstrap or bound): Y = cointegrated, N = not cointegrated, D1 = degenerate of type 1, D2 = degenerate of type 2, U=inconclusive inference.}
  \label{tab:cointbig}%
\end{table}%
\newpage

\section{Conclusion}\label{sec:5sec}
In this paper we propose a new bootstrap approach to determine the presence of cointegrating relationships in an ARDL model. The analysis here developed goes beyond the simple bivariate model used by~\citet{mcnown2018bootstrapping} and employs an approach for estimating the marginal VECM model for the regressors which ensures the independence of the related estimates from those of the ARDL equation. Monte Carlo simulations under different data generating processes, either with or without cointegrating relationships, while covering also degenerate cases, confirm that bootstrap tests perform better than the bound tests and the asymptotic $F$ test on the independent variables recently proposed by ~\citet{sam2019augmented}. Comparing the performance of the bound and bootstrap tests in both a conditional and an unconditional ARDL model proves that any inference based exclusively on the latter may lead to misleading results. 
The analysis developed in the paper focuses on two of the most frequently used specifications (restricted intercept and no trend, unrestricted intercept and no trend) of the five proposed by ~\citet{pesaran2001}.
Two illustrative applications are shown, the first focusing on German macroeconomic data, the second on the relationship between GDP, investment and exports of different OECD countries. The results highlight the drawbacks of the bound procedure, and allow to understand how adopting the unconditional specification of the ARDL model in the bootstrap approach can lead to unexpected results, potentially contradicting the more accurate analysis provided by the conditional ARDL model. General guidelines are eventually offered to effectively test for cointegration, and to avoid erroneous inference in presence of degenerate cases.
\newpage
\bibliographystyle{plainnat.bst}
\bibliography{biblio.bib}

\begin{thebibliography}{19}
\expandafter\ifx\csname natexlab\endcsname\relax\def\natexlab#1{#1}\fi
\expandafter\ifx\csname url\endcsname\relax
  \def\url#1{{\tt #1}}\fi

\bibitem[Banerjee et~al.(1993)Banerjee, Dolado, Galbraith, and
  Hendry]{banerjee1993co}
{\sc Banerjee, Anindya and Dolado, Juan J and Galbraith, John W and Hendry,
  David}.
\newblock {\em Co-integration, error correction, and the econometric analysis
  of non-stationary data}.
\newblock Oxford university press, 1993.

\bibitem[Boswijk(1995)]{boswijk1995efficient}
{\sc Boswijk, H Peter}.
\newblock Efficient inference on cointegration parameters in structural error
  correction models.
\newblock {\em Journal of Econometrics}, 69\penalty0 (1):\penalty0 133--158,
  1995.

\bibitem[Davidson et~al.(1978)Davidson, Hendry, Srba, and
  Yeo]{davidson1978econometric}
{\sc Davidson, James EH and Hendry, David F and Srba, Frank and Yeo, Stephen}.
\newblock Econometric modelling of the aggregate time-series relationship
  between consumers' expenditure and income in the united kingdom.
\newblock {\em The Economic Journal}, pages 661--692, 1978.

\bibitem[Davidson and MacKinnon(2005)]{davidson2005case}
{\sc Davidson, Russell and MacKinnon, James G}.
\newblock The case against jive.
\newblock {\em Journal of Applied Econometrics}, 21\penalty0 (6):\penalty0
  827--833, 2005.

\bibitem[Engle and Granger(1987)]{engle1987co}
{\sc Engle, Robert F and Granger, Clive WJ}.
\newblock Co-integration and error correction: representation, estimation, and
  testing.
\newblock {\em Econometrica: journal of the Econometric Society}, pages
  251--276, 1987.

\bibitem[Goh et~al.(2017)Goh, Sam, and McNown]{McNown2017}
{\sc Goh, Soo Khoon and Sam, Chung Yan and McNown, Robert}.
\newblock Re-examining foreign direct investment, exports, and economic growth
  in asian economies using a bootstrap ardl test for cointegration.
\newblock {\em Journal of Asian Economics}, 51:\penalty0 12--22, 2017.
\newblock ISSN 1049-0078.

\bibitem[Granger(1981)]{granger1981some}
{\sc Granger, Clive WJ}.
\newblock Some properties of time series data and their use in econometric
  model specification.
\newblock {\em Journal of econometrics}, 16\penalty0 (1):\penalty0 121--130,
  1981.

\bibitem[Johansen and Juselius(1990)]{johansen1990maximum}
{\sc Johansen, S{\o}ren and Juselius, Katarina}.
\newblock Maximum likelihood estimation and inference on cointegration—with
  applications to the demand for money.
\newblock {\em Oxford Bulletin of Economics and statistics}, 52\penalty0
  (2):\penalty0 169--210, 1990.

\bibitem[Kanioura and Turner(2005)]{kanioura2005critical}
{\sc Kanioura, Athina and Turner, Paul}.
\newblock Critical values for an f-test for cointegration in a multivariate
  model.
\newblock {\em Applied Economics}, 37\penalty0 (3):\penalty0 265--270, 2005.

\bibitem[Kripfganz and Schneider(2020)]{kripfganz2020response}
{\sc Kripfganz, Sebastian and Schneider, Daniel C}.
\newblock Response surface regressions for critical value bounds and
  approximate p-values in equilibrium correction models 1.
\newblock {\em Oxford Bulletin of Economics and Statistics}, 82\penalty0
  (6):\penalty0 1456--1481, 2020.

\bibitem[L{\"u}tkepohl(2005)]{lutkepohl2005}
{\sc L{\"u}tkepohl, Helmut}.
\newblock {\em New introduction to multiple time series analysis}.
\newblock Springer Science \& Business Media, 2005.

\bibitem[McNown et~al.(2018)McNown, Sam, and Goh]{mcnown2018bootstrapping}
{\sc McNown, Robert and Sam, Chung Yan and Goh, Soo Khoon}.
\newblock Bootstrapping the autoregressive distributed lag test for
  cointegration.
\newblock {\em Applied Economics}, 50\penalty0 (13):\penalty0 1509--1521, 2018.

\bibitem[Mills and Pentecost(2001)]{mills2001real}
{\sc Mills, Terence C and Pentecost, Eric J}.
\newblock The real exchange rate and the output response in four eu accession
  countries.
\newblock {\em Emerging Markets Review}, 2\penalty0 (4):\penalty0 418--430,
  2001.

\bibitem[Narayan(2005)]{narayan2005saving}
{\sc Narayan, Paresh Kumar}.
\newblock The saving and investment nexus for china: evidence from
  cointegration tests.
\newblock {\em Applied economics}, 37\penalty0 (17):\penalty0 1979--1990, 2005.

\bibitem[Narayan and Smyth(2004)]{narayan2004crime}
{\sc Narayan, Paresh Kumar and Smyth, Russell}.
\newblock Crime rates, male youth unemployment and real income in australia:
  evidence from granger causality tests.
\newblock {\em Applied Economics}, 36\penalty0 (18):\penalty0 2079--2095, 2004.

\bibitem[Narayan and Smyth(2015)]{nar2015}
{\sc Narayan, Seema and Smyth, Russell}.
\newblock The financial econometrics of price discovery and predictability.
\newblock {\em International Review of Financial Analysis}, 42:\penalty0
  380--393, 2015.

\bibitem[Pesaran et~al.(2001)Pesaran, Shin, and Smith]{pesaran2001}
{\sc Pesaran, M Hashem and Shin, Yongcheol and Smith, Richard J}.
\newblock Bounds testing approaches to the analysis of level relationships.
\newblock {\em Journal of applied econometrics}, 16\penalty0 (3):\penalty0
  289--326, 2001.

\bibitem[Rao(1997)]{rao1997cointegration}
{\sc Rao, B Bhaskara}.
\newblock {\em Cointegration for the applied economist}.
\newblock Allied Publishers, 1997.

\bibitem[Sam et~al.(2019)Sam, McNown, and Goh]{sam2019augmented}
{\sc Sam, Chung Yan and McNown, Robert and Goh, Soo Khoon}.
\newblock An augmented autoregressive distributed lag bounds test for
  cointegration.
\newblock {\em Economic Modelling}, 80:\penalty0 130--141, 2019.

\end{thebibliography}

\newpage
\appendix
\section{The Methodological Framework of Bound Tests}\label{app:Appendix}
Let us consider a VAR model specified as in \eqref{eq:1eq}. Expanding the matrix polynomial $\bm{A}(z)$ about $z=1$, yields
\begin{equation}\label{eq:28eq}
\bm{A}(z)=\bm{A}(1)+(1-z)\bm{Q}(z)
\end{equation}
where
\begin{equation}\label{eq:29eq}
\bm{A}(1)=\bm{I}_{K+1}-\sum_{j=1}^{p}\bm{A}_{j}, \enspace \enspace\bm{Q}(z)=\sum_{j=1}^{p}(-1)^{j}(1-z)^{j-1}\frac{1}{j!}\bm{A}^{(j)}(1), \enspace \enspace \bm{A}^{(j)}(1)=\frac{\partial^{j}\bm{A}(z)}{\partial z^j }_{|z=1}
\end{equation}
Formula \eqref{eq:28eq} can be rewritten as
\begin{equation}\label{eq:30eq}
\bm{A}(z)=\bm{A}(1)z+(1-z)\bm{\Gamma}(z)\end{equation}
where
\begin{equation}\label{eq:31eq}
\bm{\Gamma}(z)=(\bm{Q}(z)+\bm{A}(1))=\bm{I}_{K+1}-\sum_{i=1}^{p-1}\bm{\Gamma}_{i}z^i, \enspace \enspace \bm{\Gamma}_{i}=-\sum_{j=i+1}^{p}\bm{A}_j
\end{equation}
The VECM representation of \eqref{eq:1eq} follows accordingly, that is
\begin{equation}\label{eq:32eq}
\Delta\bm{z}_t=\bm{\alpha}_{0}+\bm{\alpha}_{1}t-\bm{A}(1)\bm{z}_{t-1}+\sum_{j=1}^{p-1}\bm{\Gamma}_{j}\Delta \bm{z}_{t-j}+\bm{\varepsilon}_t
\end{equation}
where $\Delta=\bm{I}-L$ is the backward difference operator, and
\begin{equation}\label{eq:33eq}
\bm{\alpha}_0=\bm{A}(1)\bm{\mu}+\bm{Q}(1)\bm{\eta}=\bm{A}(1)\bm{\mu}+(\bm{\Gamma}(1)-\bm{A}(1))\bm{\eta}, \enspace \enspace \enspace \bm{\alpha}_1=\bm{A}(1)\bm{\eta}
\end{equation}
The matrix $\bm{A}(1)$ is assumed to be singular, thus allowing the components of $\bm{z}_t$ to be integrated and  possibly cointegrated. Should this the case and assuming cointegration among the explanatory variables, the following holds \footnote{ If the explanatory variables are stationary $\bm{A}_{xx}$ is non-singular ($rk(\bm{A}_{xx})=K$), while when they are integrated but without cointegrating relationship $\bm{A}_{xx}$ is a null matrix }
\begin{center}
    $\bm{A}(1)=\begin{bmatrix}
\underset{(1,1)}{a_{yy}} & \underset{(1,K)}{\bm{a}_{yx}^{'}} \\ \underset{(K,1)}{\bm{a}_{xy}} & \underset{(K,K)}{\bm{A}_{xx}}  
\end{bmatrix}=\underset{(K+1,r+1)}{\bm{B}}\underset{(r+1,K+1)}{\bm{C}^{'}}=\begin{bmatrix}b_{yy} & \bm{b}_{yx}^{'}\\  \bm{b}_{xy} & \bm{B}_{xx} \end{bmatrix}\begin{bmatrix}c_{yy} & \bm{c}_{yx}^{'}\\  \bm{c}_{xy} & \bm{C}_{xx}'\end{bmatrix}=$
\end{center}
\begin{equation}\label{eq:34eq}
=\begin{bmatrix}b_{yy}c_{yy}+\bm{b}_{yx}^{'}\bm{c}_{xy} & b_{yy}\bm{c}_{yx}^{'}+\bm{b}_{yx}^{'}\bm{C}_{xx}'\\
\bm{b}_{xy}c_{yy}+\bm{B}_{xx}\bm{c}_{xy} & \bm{b}_{xy}\bm{c}_{yx}^{'}+ \bm{A}_{xx} \end{bmatrix}, \enspace \enspace \enspace rk(\bm{A}(1))=rk(\bm{B})=rk(\bm{C})
\end{equation}
where $\bm{B}$ and $\bm{C}$ are full column rank matrices arising from the rank-factorization of $\bm{A}(1)=\bm{B}\bm{C}'$ with $\bm{C}$ matrix of the long-run relationships of the process    
\begin{equation}\label{eq:35eq}
\bm{C}^{'}\bm{z}_{t-1}\sim \bm{I}(0)
\end{equation}
and $\bm{B}_{xx}$, $\bm{C}_{xx}$ arising from the rank factorization of $\bm{A}_{xx}=\bm{B}_{xx}\bm{C}_{xx}'$, with $rk(\bm{A}_{xx})=rk(\bm{B}_{xx})=rk(\bm{C}_{xx})=r$. \\
To study the adjustment to the equilibrium of a single variable $y_t$, given the other $\bm{x}_t$ variables, let us partition the vectors $\bm{z}_t$ and $\bm{\varepsilon}_t$ as follows
\begin{equation}\label{eq:36eq}
\bm{z}_t=\begin{bmatrix}
\underset{(1,1)}{y_{t}}  \\ \underset{(K,1)}{\bm{x}_{t}}  
\end{bmatrix}, \enspace \enspace \enspace \bm{\varepsilon}_t=\begin{bmatrix}
\underset{(1,1)}{\varepsilon_{yt}} \\  \underset{(K,1)}{\bm{\varepsilon}_{xt}} 
\end{bmatrix}
\end{equation}
Under the assumption 
\begin{equation}\label{eq:37eq}
\bm{\varepsilon}_t \sim N\Bigg(\bm{0}, \begin{bmatrix}
\underset{(1,1)}{\sigma_{yy}}& \underset{(1,K)}{\bm{\sigma}_{yx}^{'}}   \\ \underset{(K,1)}{\bm{\sigma}_{xy}} & \underset{(K,K)}{\bm{\Sigma}_{xx}} \end{bmatrix}\Bigg)  
\end{equation}
the following holds
\begin{equation}\label{eq:38eq}
\varepsilon_{yt}=\bm{\omega}^{'}\bm{\varepsilon}_{xt}+\nu_{yt} \sim N(0,\sigma_{y.x})      
\end{equation}
where $\sigma_{y.x}=\sigma_{yy}-\bm{\omega}^{'}\bm{\sigma}_{xy}$ with $\bm{\omega}^{'}=\bm{\sigma}^{'}_{yx}\bm{\Sigma}^{-1}_{xx}$, and $\nu_{yt}$ is independent of $\bm{\varepsilon}_{xt}$.\\
By partitioning the vectors $\bm{\alpha}_{0}$, $\bm{\alpha}_{1}$, the matrix $\bm{A}(1)$ and the polynomial matrix $\bm{\Gamma}(L)$ conformably to $\bm{z}_{t}$, as follows
\begin{equation}\label{eq:39eq}
\bm{\alpha}_0=\begin{bmatrix}
\underset{(1,1)}{\alpha_{0y}}  \\ \underset{(K,1)}{\bm{\alpha}_{0x}} 
\end{bmatrix}, \enspace \enspace \enspace \bm{\alpha}_1=\begin{bmatrix}
\underset{(1,1)}{\alpha_{1y}}  \\ \underset{(K,1)}{\bm{\alpha}_{1x} }
\end{bmatrix}
\end{equation}
\begin{equation}\label{eq:40eq}
\bm{A}(1)=\begin{bmatrix}
\underset{(1,K+1)}{\bm{a}^{'}_{(y)}}  \\ \underset{(K,K+1)}{\bm{A}_{(x)}} 
\end{bmatrix}=\begin{bmatrix}
\underset{(1,1)}{a_{yy}} & \underset{(1,K)}{\bm{a}^{'}_{yx}}  \\ \underset{(K,1)}{\bm{a}_{xy}} & \underset{(K,K)}{\bm{A}_{xx} }
\end{bmatrix}, \enspace \enspace \enspace
\bm{\Gamma}(L)=\begin{bmatrix}
\underset{(1,K+1)}{\bm{\gamma}^{'}_{y}(L)}  \\ \underset{(K,K+1)}{\bm{\Gamma}_{(x)}(L)} 
\end{bmatrix}=\begin{bmatrix}
\underset{(1,1)}{\gamma_{yy}(L)} & \underset{(1,K)}{\bm{\gamma}^{'}_{yx}(L)}  \\ \underset{(K,1)}{\bm{\gamma}_{xy}(L)} & \underset{(K,K)}{\bm{\Gamma}_{xx}(L) }
\end{bmatrix}
\end{equation}
and substituting \eqref{eq:38eq} into \eqref{eq:32eq} yields
\begin{equation}\label{eq:41eq}
\Delta\bm{z}_t=\begin{bmatrix}
\Delta y_{t} \\ \Delta\bm{x}_{t} 
\end{bmatrix}=\begin{bmatrix}
\alpha_{0.y} \\ \bm{\alpha}_{0x} 
\end{bmatrix} + \begin{bmatrix}
\alpha_{1.y} \\ \bm{\alpha}_{1x} 
\end{bmatrix}t- \begin{bmatrix}
\bm{a}^{'}_{(y).x} \\ \bm{A}_{(x)} 
\end{bmatrix}\begin{bmatrix}
y_{t-1} \\ \bm{x}_{t-1} 
\end{bmatrix} + \begin{bmatrix}
\bm{\gamma}^{'}_{y.x}(L) \\ \bm{\Gamma}_{(x)}(L) 
\end{bmatrix}\Delta\bm{z}_t+\begin{bmatrix}
\bm{\omega}^{'}\Delta\bm{x}_{t} \\ \bm{0} 
\end{bmatrix}+\begin{bmatrix}
{\nu}_{yt} \\ \bm{\varepsilon}_{xt} 
\end{bmatrix} 
\end{equation}
where 
\begin{equation}\label{eq:42eq}
\alpha_{0.y}=\alpha_{0y}-\bm{\omega}^{'}\bm{\alpha}_{0x}, \enspace \enspace \enspace \alpha_{1.y}=\alpha_{1y}-\bm{\omega}^{'}\bm{\alpha}_{1x}
\end{equation}
\begin{equation}\label{eq:43eq}
\bm{a}^{'}_{(y).x}=\bm{a}^{'}_{(y)}-\bm{\omega}^{'}\bm{A}_{(x)}, \enspace \enspace \enspace \bm{\gamma}^{'}_{y.x}(L)=\sum_{j=1}^{p-1}\bm{\gamma}^{'}_{j}L^{j}=\bm{\gamma}_{y}'(L)-\bm{\omega}'\bm{\Gamma}_{(x)}(L)
\end{equation}
Note that, according to \eqref{eq:41eq}, \eqref{eq:42eq} and  \eqref{eq:43eq}, conditioning $y_{t}$ on $\bm{x}_{t}$, modifies  intercept, slope, coefficient of the lagged values of the independent variables, short-run components of the ARDL equation 
and introduces in this equation unlagged differences of $\bm{x}_{t}$, namely $\bm{\omega}^{'}\Delta\bm{x}_t$.\\
In light of \eqref{eq:41eq}, the long-run relationships of the VECM turn out to be now included in the matrix
\begin{equation}\label{eq:44eq}
\begin{bmatrix}
\bm{a}^{'}_{(y).x} \\ \bm{A}_{(x)} 
\end{bmatrix}=\begin{bmatrix}
a_{yy}-\bm{\omega}^{'}\bm{a}_{xy} & \bm{a}_{yx}^{'}-\bm{\omega}^{'}\bm{A}_{xx} \\ \bm{a}_{xy}&\bm{A}_{xx} 
\end{bmatrix}=\bm{B}_{y.x}\bm{C}'
\end{equation}
where (see \cite{boswijk1995efficient})
\begin{equation}
\bm{B}_{y.x}=\begin{bmatrix}b_{yy}-\bm{w}'\bm{b}_{xy} , & \bm{b}_{yx}'-\bm{w}'\bm{B}_{xx}\\
\bm{b}_{xy} , & \bm{B}_{xx}
\end{bmatrix}
=\begin{bmatrix}\bm{b}_{y.x}'\\ \bm{B}_{(x)}
\end{bmatrix}
\end{equation}
and accordingly, the conditional and the marginal model become
\begin{align}\label{eq:mar1}
\Delta y_t&= \alpha_{0y}+ \alpha_{1y}t+\bm{b}_{y.x}'\bm{C}'\bm{z}_{t-1}+\bm{\gamma}^{'}_{y.x}(L)\Delta\bm{z}_{t}+\bm{\omega}^{'}\Delta\bm{x}_{t}+\nu_{yt}\\\label{eq:con1}
\Delta \bm{x}_t&= \alpha_{0x}+ \alpha_{1x}t+\bm{B}_{(x)}'\bm{C}'\bm{z}_{t-1}+\bm{\Gamma}^{'}_{(x)}(L)\Delta\bm{z}_{t}+\epsilon_{xt}
\end{align}
Looking at \eqref{eq:mar1} and \eqref{eq:con1} we see that the cointegration relationships between $y_{t}$ and $\bm{x}_{t}$ appear both in the conditional and the marginal models.  
To rule out the presence of long-run relationships between $y_{t}$ and $\bm{x}_{t}$ in the marginal model, 
$\bm{a}_{xy}$ is assumed to be a null vector (and, accordingly, $\bm{b}_{xy}$ and $\bm{c}_{xy}$). This assumption, together with the operation of conditioning $y_{t}$ on $\bm{x}_{t}$, makes the $\bm{x}_{t}$ variables weakly exogenous with respect to the cointegrating vector included in the ARDL equation, thus allowing to rule out the marginal model in the analysis of the cointegration between $y_{t}$ and $\bm{x}_{t}$. Under this assumption, the marginal model becomes
\begin{equation}\label{eq:con2}
\Delta \bm{x}_t= \alpha_{0x}+ \alpha_{1x}t+\bm{B}_{xx}\bm{C}_{xx}'\bm{z}_{t-1}+\bm{\Gamma}^{'}_{(x)}(L)\Delta\bm{z}_{t}+\epsilon_{xt}
\end{equation}
and turns out to include only the cointegrating relationships between the $\bm{x}_{t}$ variables. \\
As for \eqref{eq:44eq}, it becomes
\begin{equation}\label{eq:cond}
\widetilde{\bm{A}}=\begin{bmatrix}a_{yy} & \bm{a}^{'}_{yx}-\bm{\omega}^{'}\bm{A}_{xx} \\ \bm{0} & \bm{A}_{xx} 
\end{bmatrix}=\begin{bmatrix}
a_{yy} & \widetilde{\bm{a}}_{y.x}^{'} \\ \bm{0}&\bm{A}_{xx}\end{bmatrix}
\end{equation}
and can be factored as follows
\begin{equation}\label{eq:45eq}
\begin{bmatrix}
b_{yy}c_{yy} & b_{yy}\bm c_{yx}^{'}+(\bm{b}_{yx}^{'}-\bm{\omega}^{'}\bm{B}_{xx})\bm{C}_{xx}' \\ \bm{0}& \bm{B}_{xx}\bm{C}_{xx}'\end{bmatrix}
\end{equation}
In light of \eqref{eq:45eq}, the cointegrating vectors between $y_t$ and $\bm{x}_t$, namely ($c_{yy}$, $\bm{c}^{'}_{yx}$), turns out to be included only in the conditional model for $\Delta y_{t}$. \footnote{The matrix $\bm{C}_{xx}$ is the cointegrating matrix for the variables $\bm{x}_t$.}\\
\cite{pesaran2001} proved that
\begin{equation}
\widetilde{\bm{a}}_{y.x}^{'}(\bm{z}_{t}
-\bm{\mu}-\bm{\eta}t)=\nu_{t}
\end{equation}
where $\nu_{t}$ is a zero mean stationary process. This implies the existence of a conditional level relationship between $y_t$ and $\bm{x}_t$
\begin{equation}
y_{t}=\theta_{0}+\theta_{1}t+\bm{\theta}'\bm{x}_{t}+\zeta_{t}
\end{equation}
where $\bm{\theta}=\frac{\widetilde{\bm{a}}_{y.x}}{a_{yy}}$ and $\zeta_{t}$ is a zero mean stationary process.\\ Taking into account \eqref{eq:45eq}, the cointegrating vector can be also expressed as 
\begin{equation}\label{eq:con}
\bm{\theta}'=-\frac{1}{a_{yy}}\underset{(1,r+1)}{[b_{yy}, \enspace (\bm{b}_{yx}-\bm{\omega}'\bm{B}_{xx})]}
\underset{(r+1,K)}{\begin{bmatrix} \bm{c}'_{yx}\\ \bm{C}'_{xx} \end{bmatrix}}=\underset{(1,r+1)}{\bm{g}'}\underset{(r+1,K)}{\begin{bmatrix} \bm{c}'_{yx}\\ \bm{C}'_{xx} \end{bmatrix}}
\end{equation}
Formula \eqref{eq:con} leads to the following conclusion
\begin{equation}\label{eq:rank}
rk\begin{bmatrix}\bm{c}'_{yx}\\ \bm{C}'_{xx}\end{bmatrix}=\begin{cases}
r  \to \enspace y_{t} \sim I(0) \\
r+1 \to \enspace y_{t} \sim I(1) 
\end{cases}
\end{equation}
In order to get a deeper insight into \eqref{eq:rank}, note that the following holds for the cointegrating matrix $\widetilde{\bm{A}}$ 
\begin{equation}\label{eq:46eq}
r\leq rk(\widetilde{\bm{A}}) \leq r+1
\end{equation}
where $r=rk(\bm{A}_{xx})$ and $0 \leq r\leq K$.  \\
In particular, $rk(\widetilde{\bm{A}})=r$ when $a_{yy}=0$. If this is the case,  each of the following two conditions 
\begin{align}\label{eq:47eq}
\nonumber
\widetilde{\bm{a}}_{y.x}&=\bm{0}\\
\text{or}\\
\widetilde{\bm{a}}_{y.x} &\neq \bm{0}\nonumber
\end{align}
is worth considering. In the former case, no level relationship between $y_{t}$ and $\bm{x}_{t}$ can exist in the ARDL equation given by the first equation of \eqref{eq:41eq}. 
In the latter case, the said equations turns out include only the one-lagged cointegrating relationships of the explanatory variables
\begin{equation}
\Delta y_{t}=\alpha_{0.y}+\alpha_{1.y}t-(\bm{b}_{yx}^{'}-\bm{\omega}^{'}\bm{B}_{xx})\bm{C}_{xx}'\bm{x}_{t-1}+ \bm{\gamma}^{'}_{y.x}(L)\Delta\bm{z}_{t} +\bm{\omega}^{'}\Delta\bm{x}_{t} +{\nu}_{yt}
\end{equation}
Thus, we conclude that $y_t$ must be first order integrated without a long-run relationship with $\bm{x}_t$. \\ 
Now, let $rk(\widetilde{\bm{A}})=rk(\bm{A}_{xx})+1$. Here the following two cases are worth analyzing
\begin{equation}\label{eq:50eq}
\left\{\begin{array}{ll}
1) \enspace \widetilde{\bm{a}}'_{y.x}\neq\bm{0}' \to \begin{cases}
1_{a}) \enspace \widetilde{\bm{a}}'_{y.x}=\bm{\omega}'\bm{A}_{xx}\\
1_{b}) \enspace \widetilde{\bm{a}}'_{y.x}=\bm{a}'_{yx}
\end{cases}\\
\\
2) \enspace \widetilde{\bm{a}}'_{y.x}=\bm{0}' \\
\end{array}\right.
\end{equation}
When both $a_{yy}$ and $\widetilde{\bm{a}}^{'}_{y.x}$ are not null, cointegrating relationships may be present in the ARDL equation. 
Bearing in mind that $\widetilde{\bm{a}}'_{y.x}=\bm{a}_{yx}'-\bm{w}\bm{A}_{xx}$, it is easy to see that when $\bm{a}'_{yx}=\bm{0}'$ the following holds
\begin{equation}\label{eq:1a}
\bm{\theta}'=-\frac{\bm{\omega}'\bm{A}_{xx}}{a_{yy}}=\frac{1}{a_{yy}} \begin{bmatrix} b_{yy}, \enspace -\bm{\omega}'\bm{B}_{xx} \end{bmatrix}
\begin{bmatrix}0 \\\bm{C}_{xx}'\end{bmatrix}, \enspace rk
\begin{bmatrix}0 \\\bm{C}_{xx}'\end{bmatrix}=r
\end{equation}
Thus, $y_{t}$  must be stationary (or trend stationary depending on $\alpha_{1y}$ is null or not), for whatever value of $r$.\\ In case $1_{b})$ the explanatory variables are integrated without cointegrating relationship, and the long-run relationship between $y_{t}$ and $\bm{x}_{t}$ turns out to be
\begin{equation}\label{eq:CC}
\bm{\theta}'=\frac{\bm{a}'_{yx}}{a_{yy}}= \frac{1}{a_{yy}}\begin{bmatrix} b_{yy}, \enspace \bm{b}'_{yx} \end{bmatrix}
\begin{bmatrix} \bm{c}'_{yx} \\ \bm{0} \end{bmatrix}, \enspace rk
\begin{bmatrix} \bm{c}'_{yx} \\ \bm{0} \end{bmatrix}=r+1, \enspace r=0
\end{equation}
 Thus, $y_{t}$ must be cointegrated with the explanatory variables.\\
Finally, if $\widetilde{\bm{a}}_{y.x}=\bm{0}$, as assumed by case 2), then a long-run relationship between $y_{t}$ and $\bm{x}_{t}$ cannot exist. Accordingly,$y_{t}$ must be stationary or trend stationary.
\\
\\
~\citet{pesaran2001} introduced five different specifications for the ARDL model 
\begin{equation}\label{eq:ARDL}
\Delta y_t=\alpha_{0.y}+\alpha_{1.y}t-a_{yy}y_{t-1}-\widetilde{\bm{a}}^{'}_{y.x}\bm{x}_{t-1}+\sum_{j=1}^{p-1}\bm{\gamma}^{'}_{y.x,j}\Delta\bm{z}_{t-j}+\bm{\omega}^{'}\Delta\bm{x}_{t}+\nu_{yt}
\end{equation}
which depend on the specifications of the deterministic components. Before introducing these specifications, let us note that in light of \eqref{eq:33eq} the drift and the trend coefficient in the conditional VECM \eqref{eq:41eq} turn out to be defined as 
\begin{equation}
\bm{\alpha_{0}}^{c}=\widetilde{\bm{A}}(1)(\bm{\mu}-\bm{\eta})+\widetilde{\bm{\Gamma}}(1)\bm{\eta}, \enspace \enspace \bm{\alpha_{1}}^{c}=\widetilde{\bm{A}}(1)\bm{\eta}
\end{equation}
where $\widetilde{\bm{A}}(1)$ is as in \eqref{eq:cond} and $\widetilde{\bm{\Gamma}}(1)=\begin{bmatrix} \bm{\gamma}_{y.x}'(1) \\ \bm{\Gamma}_{(x)}(1) \end{bmatrix}$. Accordingly, 
\begin{equation}
\alpha_{0.y}= \bm{e}_{1}'\bm{\alpha_{0}}^{c}=\begin{bmatrix} a_{yy} ,\enspace \widetilde{\bm{a}}'_{y.x}
\end{bmatrix}(\bm{\mu}-\bm{\eta})+\bm{\gamma}'_{y.x}(1)\bm{\eta}, \enspace \enspace \alpha_{1.y}=\bm{e}_{1}'\bm{\alpha_{1}}^{c}=\begin{bmatrix}a_{yy},\enspace \widetilde{\bm{a}}'_{y.x}\end{bmatrix}\bm{\eta} 
\end{equation}
where $\bm{e}_{1}$ is the $(K+1)$ first elementary vector.\\ 
Then, possible specifications for the ARDL equation \eqref{eq:ARDL} are the following 
\begin{enumerate}
\item \textit{No intercept and no trend} $(\bm{\mu}=\bm{\eta}=\bm{0})$.\\This entails $\alpha_{0.y}=\alpha_{1.y}=0$. Accordingly
\begin{align}\label{eq:57eq}
\Delta y_t&=-a_{yy}y_{t-1}-\widetilde{\bm{a}}^{'}_{y.x}\bm{x}_{t-1}+\sum_{j=1}^{p-1}\bm{\gamma}^{'}_{y.x,j}\Delta\bm{z}_{t-j}+\bm{\omega}^{'}\Delta\bm{x}_{t}+\nu_{yt}\\
&=-a_{yy}EC_{t-1}+\sum_{j=1}^{p-1}\bm{\gamma}^{'}_{y.x,j}\Delta\bm{z}_{t-j}+\bm{\omega}^{'}\Delta\bm{x}_{t}+\nu_{yt}\nonumber
\end{align}  
where $EC_{t-1}=y_{t-1}-\bm{\delta}'_{y
.x}\bm{x}_{t-1}$, with $\bm{\delta}'_{y
.x}=-\frac{{\widetilde{\bm{a}}}'_{y.x}}{a_{yy}}$.\\
\item \textit{Restricted intercept and no trend} ($\bm{\alpha}_{0}^{c}=\widetilde{\bm{A}}(1)\bm{\mu}$, $\bm{\eta}=\bm{0}$).\\This entails $\alpha_{0.y}=\begin{bmatrix} a_{yy}, & \widetilde{\bm{a}}'_{y.x}\end{bmatrix}\begin{bmatrix}
\underset{(1,1)}{\mu_{y}}\\ \underset{(K,1)}{\bm{\mu}_{x}} 
\end{bmatrix} =a_{yy}\mu_{y}+\widetilde{\bm{a}}'_{y.x}\bm{\mu}_{x}$ and $\alpha_{1.y}=0$. Accordingly, demeaned variables appear in the ARDL equation
\begin{align}\label{eq:58eq}
\Delta y_{t}&=-a_{yy}(y_{t-1}-\mu_{y})-\widetilde{\bm{a}}^{'}_{y.x}(\bm{x}_{t-1}-\bm{\mu}_{x})+\sum_{j=1}^{p-1}\bm{\gamma}^{'}_{y.x,j}\Delta\bm{z}_{t-j}+\bm{\omega}^{'}\Delta\bm{x}_{t}+\nu_{yt}\\
&=-a_{yy}EC_{t-1}+\sum_{j=1}^{p-1}\bm{\gamma}^{'}_{y.x,j}\Delta\bm{z}_{t-j}+\bm{\omega}^{'}\Delta\bm{x}_{t}+\nu_{yt}\nonumber
\end{align} 
where $EC_{t-1}=y_{t-1}-\delta_{0}-\bm{\delta}'_{y.x}\bm{x}_{t-1}$ with $\bm{\delta}'_{y.x}=-\frac{\widetilde{\bm{a}}^{'}_{y.x}}{a_{yy}}$ and $\bm{\delta}_{0}=\mu_{y}-\bm{\delta}'_{y.x}\bm{\mu}_{x}$.
\item \textit{Unrestricted intercept and no trend} ($\bm{\alpha}_{0}^{c}\neq\widetilde{\bm{A}}(1)\bm{\mu}$,  $\bm{\eta}=\bm{0}$).\\
Thus, $\alpha_{0.y} \neq 0$, $\alpha_{1.y} = 0$. Accordingly, an intercept is included in the ARDL equation
\begin{align}\label{eq:59eq}
\Delta y_{t}&=\alpha_{0.y}-a_{yy}y_{t-1}-
\widetilde{\bm{a}}^{'}_{y.x}\bm{x}_{t-1}+\sum_{j=1}^{p-1}\bm{\gamma}^{'}_{y.x,j}\Delta\bm{z}_{t-j}+\bm{\omega}^{'}\Delta\bm{x}_{t}+\nu_{yt}\\
&=\alpha_{0y}-a_{yy}EC_{t-1}+\sum_{j=1}^{p-1}\bm{\gamma}^{'}_{y.x,j}\Delta\bm{z}_{t-j}+\bm{\omega}^{'}\Delta\bm{x}_{t}+\nu_{yt}\nonumber
\end{align}  
where $EC_{t-1}=y_{t-1}-\bm{\delta}_{y.x}'\bm{x}_{t-1}$, with $\bm{\delta}_{y.x}'=-\frac{\widetilde{\bm{a}}^{'}_{y.x}}{a_{yy}}$. Note that the presence of a drift in the above equation implies the presence of a linear trend in the original series $y_{t}$.\footnote{In this case, to have cointegrating relationships with null mean, the EC term is sometimes reformulated as follows 
\begin{equation*}
EC_{t-1}=y_{t-1}-\bm{\delta}_{y.x}'\bm{x}_{t-1}-\widehat{\mu}_{EC}
\end{equation*}
where $\widehat{\mu}_{EC}=\frac{1}{T}\sum_{t=2}^{T}EC_{t-1}$ and $T$ is the sample size. Then, the ARDL equation is reformulated as follows
\begin{equation*}
\Delta y_{t}=\alpha^{*}_{0y}-a_{yy}EC_{t-1}+\sum_{j=1}^{p-1}\bm{\gamma}^{'}_{y.x,j}\Delta \bm{z}_{t-j}+\bm{\omega}'\Delta \bm{x}_{t}+\nu_{yt}
\end{equation*}
where $\alpha^{*}_{0y}=\alpha_{0y}+a_{yy}\widehat{\mu}_{EC}$.} 
\item \textit{Unrestricted intercept, restricted trend} ($\bm{\alpha_{0}}^{c}\neq\widetilde{\bm{A}}(1)(\bm{\mu}-\bm{\eta})+\widetilde{\bm{\Gamma}}(1)\bm{\eta}$, $ {\bm{\alpha}}_{1}^{c}=\widetilde{\bm{A}}(1)\bm{\eta}$).\\ Hence, $\alpha_{0.y}\neq 0$,
$\alpha_{1.y}=\begin{bmatrix}a_{yy}, \widetilde{\bm{a}}^{'}_{y.x}\end{bmatrix}\begin{bmatrix}
\underset{(1,1)}{\eta_{y} }\\ \underset{(K,1)}{\bm{\eta}_{x}}
\end{bmatrix}=a_{yy}\eta_{y}+\widetilde{\bm{a}}_{y.x}\bm{\eta}_{x}$. Accordingly 
\begin{align}\label{eq:60eq}
\Delta y_{t}&=\alpha_{0.y}
-a_{yy}(y_{t-1}-\eta_{y}t)-\widetilde{\bm{a}}^{'}_{y.x}(\bm{x}_{t-1}-\bm{\eta}_{x}t)+\sum_{j=1}^{p-1}\bm{\gamma}^{'}_{y.x,j}\Delta\bm{z}_{t-j}+\bm{\omega}^{'}\Delta\bm{x}_{t}+\nu_{yt}\\
&=\alpha_{0.y}-a_{yy}EC_{t-1}+\sum_{j=1}^{p-1}\bm{\gamma}^{'}_{y.x,j}\Delta\bm{z}_{t-j}+\bm{\omega}^{'}\Delta\bm{x}_{t}+\nu_{yt}\nonumber
\end{align}
where $EC_{t-1}=y_{t-1}-\eta_{0}t-\bm{\delta}'_{y.x}\bm{x}_{t-1}$ with $\bm{\delta}'_{y.x}=-\frac{\widetilde{\bm{a}}^{'}_{y.x}}{a_{yy}}$ and $\eta_{0}=\eta_{y}-\bm{\delta}'_{y.x}\bm{\eta}_{x}$.
\item \textit{Unrestricted intercept, unrestricted trend} ($\bm{\alpha_{0}}^{c}\neq\widetilde{\bm{A}}(1)(\bm{\mu}-\bm{\eta})+\widetilde{\bm{\Gamma}}(1)\bm{\eta}$, $ {\bm{\alpha}}_{1}^{c}\neq\widetilde{\bm{A}}(1)\bm{\eta}$).\\ Hence, $\alpha_{0.y} \neq 0$, $\alpha_{1.y} \neq 0$. Accordingly
\begin{align}\label{eq:60eq}
\Delta y_{t}&=\alpha_{0.y}+\alpha_{1.y}t -a_{yy}y_{t-1}-\widetilde{\bm{a}}^{'}_{y.x}\bm{x}_{t-1}+\sum_{j=1}^{p-1}\bm{\gamma}^{'}_{y.x,j}\Delta\bm{z}_{t-j}+\bm{\omega}^{'}\Delta\bm{x}_{t}+\nu_{yt}\\
&=\alpha_{0.y}+\alpha_{1.y}t
-a_{yy}EC_{t-1}+\sum_{j=1}^{p-1}\bm{\gamma}^{'}_{y.x,j}\Delta\bm{z}_{t-j}+\bm{\omega}^{'}\Delta\bm{x}_{t}+\nu_{yt}\nonumber
\end{align}
where $EC_{t-1}=y_{t-1}-\bm{\delta}'_{y.x}\bm{x}_{t-1}$ with $\bm{\delta}'_{y.x}=-\frac{\widetilde{\bm{a}}^{'}_{y.x}}{a_{yy}}$. Note that the presence of a linear trend in \eqref{eq:60eq}
implies the presence of a quadratic trend in the original series $y_{t}$.
\end{enumerate}
In all these cases, the deterministic components (intercept and trend coefficient) taken into account by the $F_{ov}$ test are those included in the EC term. Thus, the $F_{ov}$ test verifies the presence of a constant term in case II, and not in case III. Similarly, it tests the presence of both the trend in case IV, and not in case V. If the null of the $F_{ov}$ test is rejected, the error-correction form of the ARDL equation is estimated and the significance of the coefficient associated to the EC term is evaluated with a $t$ test.

\end{document}